\documentclass[reprint,superscriptaddress,amsmath,amssymb,aps,prb,longbibliography,floatfix]{revtex4-2}
\usepackage[utf8]{inputenc}
\usepackage{dcolumn}
\usepackage{bm}
\usepackage{graphicx}
\usepackage{amsmath}
\usepackage{amssymb}
\usepackage{placeins}
\usepackage{float}
\usepackage{color, xcolor}
\usepackage[normalem]{ulem}
\usepackage{newunicodechar}
\newunicodechar{）}{)}
\usepackage{soul}
\soulregister\cite7
\soulregister\ref7

\begin{document}
\raggedbottom
\makeatletter
\setlength{\@dblfptop}{0pt}
\setlength{\@dblfpsep}{12pt plus 4pt}
\setlength{\@dblfpbot}{0pt plus 1fil}
\makeatother
\title{Competing magnetic correlations in the square-lattice Hubbard model with spin-dependent nearest-neighbor and next-nearest-neighbor hopping}
\author{Rana Imran Mushtaq}
\thanks{These authors contributed equally.}
\affiliation{School of Science, Harbin Institute of Technology, Shenzhen, 518055, China}
\affiliation{Shenzhen Key Laboratory of Advanced Functional Carbon Materials Research and Comprehensive Application, Shenzhen 518055, China.}

\author{Ji Liu}
\thanks{These authors contributed equally.}
\affiliation{School of Science, Harbin Institute of Technology, Shenzhen, 518055, China}
\affiliation{Shenzhen Key Laboratory of Advanced Functional Carbon Materials Research and Comprehensive Application, Shenzhen 518055, China.}

\author{Yinlong Li}
\affiliation{School of Science, Harbin Institute of Technology, Shenzhen, 518055, China}
\affiliation{Shenzhen Key Laboratory of Advanced Functional Carbon Materials Research and Comprehensive Application, Shenzhen 518055, 
China.}

\author{Wing Chi Yu}
\email{wingcyu@cityu.edu.hk}
\affiliation{Department of Physics, City University of Hong Kong, Kowloon, Hong Kong
}

\author{Xiaosen Yang}
\email{yangxs@ujs.edu.cn}
\affiliation{Department of Physics, Jiangsu University, Zhenjiang, 212013, China.}

\author{Ho-Kin Tang}
\email{denghaojian@hit.edu.cn}
\affiliation{School of Science, Harbin Institute of Technology, Shenzhen, 518055, China}
\affiliation{Shenzhen Key Laboratory of Advanced Functional Carbon Materials Research and Comprehensive Application, Shenzhen 518055, China.}

\begin{abstract}
Altermagnets have spin-split bands without net magnetization. The splitting can come from a spin-dependent nearest-neighbor (NN) hopping anisotropy $t_A$ or from a spin-dependent next-nearest-neighbor (NNN) hopping $t'$. Both change sign under a $90^\circ$ rotation and are present together when the lattice has no mirror planes, yet they have so far been studied only separately. Here we study the two-dimensional square-lattice Hubbard model with both $t_A$ and $t'$, using constrained-path quantum Monte Carlo at $U=4t$ and scanning the electron filling $n$, $t_A$ and $t'$. We find four types of magnetic correlations: N\'{e}el, stripe, diagonal, and spiral spin-density-wave. The phase diagram depends on both hoppings together. Diagonal correlations require finite $t_A$ and are suppressed by $t'$. Stripe correlations appear only once $t'$ is large enough to overcome the diagonal tendency set by $t_A$. N\'eel correlations survive at large $t_A$ and large $t'$, where the NNN model alone gives stripe or spiral correlations. The noninteracting susceptibility matches the dominant wavevector at small $t'$, but not the N\'eel correlations at $t'\geq0.3$, which come from the interaction. Finite-size scaling with three lattice sizes is consistent with long-range order only near half filling; at the stripe, spiral and diagonal points we tested, the structure factor per site extrapolates to zero. The joint behavior persists up to $U=6t$.
\end{abstract}

\maketitle

\section{INTRODUCTION}
\label{Introduction}

Magnetic correlations in strongly correlated electron systems are highly sensitive to the geometry of the Fermi surface and the underlying single-particle band structure~\cite{Dagotto1994,Lee2006,Scalapino2012}. The two-dimensional Hubbard model on the square-lattice provides a standard framework for studying this problem, capturing the fundamental interplay between kinetic energy and on-site Coulomb repulsion within a minimal yet nontrivial setting~\cite{Hubbard1963,LeBlanc2015,Qin2022,Qin2020}. In this model, the preferred magnetic ordering vector can be shifted across qualitatively different regimes and the dominating particle-hole scattering channels can be redirected by variations in carrier density, hopping amplitudes, or interaction strength~\cite{Zheng2017,Arovas2022,Huang2017}. The momentum-resolved spin structure factor $S^z(\mathbf{k})$ serves as a direct diagnostic of this competition, which encodes the strength and the wavevector of spin-density-wave (SDW) correlations. How modifications to the band structure steer this competition across a broad parameter range has remained a central problem in correlated square-lattice physics.

The concept of altermagnetism, which is defined as a distinctive magnetic phase exhibiting zero net magnetization accompanied by a significant momentum-dependent spin splitting, has led to a new understanding of magnetism in correlated systems~\cite{Smejkal2022a,Smejkal2022b,Hayami2019,Yuan2020,Yuan2021,Smejkal2022GMR,McClarty2024}. Unlike conventional antiferromagnets, where a combination of time-reversal symmetry and spatial translation enforces spin degeneracy throughout the Brillouin zone, altermagnets break this equivalence through rotational symmetry operations, generating alternating-sign spin splitting whose angular pattern such as $d$-wave, $g$-wave, or higher is dictated by the rotational symmetry of the underlying lattice~\cite{Smejkal2020,Gonzalez2021,Hayami2020,Liu2022SpinGroup}. This mechanism produces highly anisotropic Fermi surfaces with opposite spin polarizations in different momentum-space sectors, even though the material carries no net moment. Experimental signatures consistent with this picture have now been reported in several candidate materials, including RuO$_2$, MnTe, and CrSb, through angle-resolved photoemission spectroscopy and related probes~\cite{Fedchenko2024,Krempasky2024,Osumi2024,Lee2024,Feng2022,Reimers2024,Ding2024,Mazin2023}. Interest has since extended from the band-structure level to correlated altermagnets. Altermagnetic order has been proposed in Fermi--Hubbard models realizable with ultracold atoms~\cite{Das2024}. At strong coupling the band splitting has been studied through altermagnetic polarons~\cite{Daghofer2026} and through the photoemission response of altermagnetic Mott insulators~\cite{Lanzini2026}. Correlated altermagnets can host Dirac points and topological phases~\cite{DelRe2025}, and the associated band crossings in two dimensions have been classified by symmetry~\cite{Parshukov2025}. Spin transport in altermagnetic Mott insulators has been proposed as an optically tunable probe~\cite{sicheler2025opticallytunablespintransport}. These works motivate a many-body study of altermagnetic hopping in an interacting lattice model, which is the subject of this work. At the level of minimal lattice models, spin-dependent hopping provides a transparent single-particle mechanism for producing this momentum-space spin splitting while keeping the total magnetization strictly zero, making such models a natural starting point for studying how altermagnetic band reconstruction modifies the magnetic response of an interacting system~\cite{Bai2024,Naka2019,Das2024,Leeb2024,Roig2024}.

Spin-dependent nearest-neighbor (NN) hopping anisotropy $t_A$ offers a minimal route to altermagnetic spin splitting in the square-lattice Hubbard model. In this setting, finite $t_A$ lifts the spin degeneracy of the van Hove singularities while keeping their momenta pinned at $(\pi, 0)$ and $(0, \pi)$, reorganizing the low-energy particle-hole phase space in a filling- and anisotropy-dependent way. The resulting interacting phase diagram reveals a three-sector structure and includes a robust high-filling N\'{e}el regime at $\mathbf{Q} = (\pi, \pi)$, a spiral sector with $\mathbf{Q} = (\pi, q)$ or $(q, \pi)$ that dominates at weak anisotropy and intermediate filling, and a diagonal sector at $\mathbf{Q} = (q, q)$ that expands progressively as $t_A$ grows stronger, as we establish below. Rather than suppressing the conventional N\'{e}el tendency, $t_A$ redirects the dominant magnetic weight into distinct momentum-space channels, with the redirection set by both filling and anisotropy. In a closely related setting, we recently showed that a spin-dependent NNN hopping $t'$ generates a $d$-wave altermagnetic splitting and drives an evolution from N\'{e}el through spiral to stripe correlations with increasing anisotropy or doping~\cite{Mushtaq2026}. Separately, the effect of $t'$ in the square-lattice Hubbard model has received extensive attention in the context of cuprate phenomenology and frustrated magnetism~\cite{Tocchio2014,Misawa2014,Moreo1991,Jiang2019}. In the conventional $t$--$t'$ Hubbard model, $t'$ lifts particle-hole symmetry, shifts the van Hove singularities away from the zone boundary $X$ points into the Brillouin zone interior, and modifies the Fermi surface in a way that perfect $(\pi, \pi)$ nesting is weakened and stripe or spiral correlations are favored at appropriate carrier densities~\cite{Tocchio2014,Misawa2014, Igoshev2010}. Despite the physical relevance of both hopping channels, the combined model in which spin-dependent anisotropy $t_A$ and spin-dependent $t'$ are present simultaneously has not been studied. It is not obvious how the two terms interact. These are qualitatively different reconstructions of the low-energy band structure, and their combined effect on the interacting magnetic phase diagram is an open question.

To address this question, we study the two-dimensional square-lattice Hubbard model in which spin-dependent NN hopping anisotropy $t_A$ and next-nearest-neighbor (NNN) hopping $t'$ are simultaneously active. Employing the constrained-path quantum Monte Carlo (CPQMC) method~\cite{Zhang1995,Zhang1997}, we evaluate the momentum-resolved spin structure factor $S^z(\mathbf{k})$ and the real-space SDW correlator $C_\mathrm{SDW}(\mathbf{r})$ across a broad three-dimensional parameter space spanned by filling $n$, anisotropy $t_A$, and NNN hopping $t'$, at interaction strength $U = 4t$. This systematic scan allows us to construct a detailed magnetic phase diagram in which the dominant ordering vector $\mathbf{Q}$ is tracked continuously as each parameter is varied. We identify four distinct magnetic correlations: the N\'{e}el at $\mathbf{Q} = (\pi, \pi)$, a stripe at $\mathbf{Q} = (\pi, 0)/(0, \pi)$ driven by directional Fermi surface nesting, a diagonal at $\mathbf{Q} = (q, q)$, and a spiral at $\mathbf{Q} = (\pi, q)/(q, \pi)$ that occupies the boundary between competing ordering tendencies. At small $t'$ the nesting of the spin-split Fermi surfaces explains the dominant wavevectors of the many-body calculation, but it does not explain the N\'eel correlations at larger $t'$, which come from the interaction. The central finding of this work is that $t_A$, $t'$, and $n$ do not act as independent control parameters. Instead, the phase boundaries depend on the parameters jointly: the diagonal correlations require both finite $t_A$ and small $t'$, and the stripe correlations appear only when $t'$ is large enough to overcome the $t_A$-driven diagonal tendency. A third joint feature is that N\'eel correlations survive at large $t_A$ and large $t'$, where the NNN model alone gives stripe or spiral correlations.

The remainder of this paper is organized as follows. Section II introduces the model and summarizes the CPQMC method. Section III analyzes the noninteracting electronic structure. Section IV presents the central magnetic results and detailed analysis. We summarize our findings and discuss their broader implications for correlated square-lattice systems in Section V.

\section{Model and Method}
\label{Model and Method}
\subsection{Model Hamiltonian}
\label{Combined model}
We study a two-dimensional square lattice Hubbard model with a NN hopping anisotropy $t_A$, and a NNN diagonal hopping $t'$. The Hamiltonian is written as $\hat{H} = \hat{H}_\text{NN} + \hat{H}_\text{NNN}$, where

\begin{equation}
\begin{split}
\hat{H}_\text{NN} = &-\sum_{i,\sigma}\Bigl(
t_{\hat{x},\sigma}\,\hat{c}_{i,\sigma}^{\dagger}\hat{c}_{i+\hat{x},\sigma}
+ t_{\hat{y},\sigma}\,\hat{c}_{i,\sigma}^{\dagger}\hat{c}_{i+\hat{y},\sigma}
+ \mathrm{H.c.}\Bigr) \\
&+ U\sum_{i}\hat{n}_{i,\uparrow}\hat{n}_{i,\downarrow}
- \mu\sum_{i,\sigma}\hat{n}_{i,\sigma},
\raisetag{1.7\baselineskip}
\end{split}
\label{eq:H_NN}
\end{equation}

\begin{equation}
\begin{split}
\hat{H}_\text{NNN} = -\sum_{i,\sigma}\Bigl(
&t'_{\hat{x}+\hat{y},\sigma}\,\hat{c}_{i,\sigma}^{\dagger}\hat{c}_{i+\hat{x}+\hat{y},\sigma}
+ t'_{\hat{x}-\hat{y},\sigma}\,\hat{c}_{i,\sigma}^{\dagger}\hat{c}_{i+\hat{x}-\hat{y},\sigma} \\
&+ \mathrm{H.c.}\Bigr).
\raisetag{1.7\baselineskip}
\end{split}
\label{eq:H_NNN}
\end{equation}
$\hat{H}_\text{NN}$ is the base NN altermagnetic Hubbard model, where $\hat{c}_{i,\sigma}^{\dagger}$ ($\hat{c}_{i,\sigma}$) creates (annihilates) an electron with spin $\sigma \in \{\uparrow, \downarrow\}$ at site $i$, and $\hat{n}_{i,\sigma} = \hat{c}_{i,\sigma}^{\dagger}\hat{c}_{i,\sigma}$ is the local number operator. The hopping anisotropy is encoded by a single tunable parameter $t_A$, with $t_{\hat{x},\uparrow} = t_{\hat{y},\downarrow} = t - t_A$ and $t_{\hat{y},\uparrow} = t_{\hat{x},\downarrow} = t + t_A$, so the two spin species experience opposite hopping anisotropy. At $t_A = 0$, $\hat{H}_\text{NN}$ reduces to the isotropic square-lattice Hubbard model. $\hat{H}_\text{NNN}$ adds the NNN hopping $t'$ with opposite signs for the two spin species, where $t'_{\hat{x}+\hat{y},\uparrow} = t'_{\hat{x}-\hat{y},\downarrow} = +t'$ and $t'_{\hat{x}-\hat{y},\uparrow} = t'_{\hat{x}+\hat{y},\downarrow} = -t'$, 
consistent with the altermagnetic symmetry of the lattice~\cite{Smejkal2022a, Smejkal2022b}. The on-site repulsion $U > 0$ drives the magnetic correlations, and $\mu$ is the chemical potential. We set $t = 1$ as the unit of energy, fix $U = 4t$, and vary $t_A$ and $t'$ independently: $t_A=0.1,0.2,\ldots,0.9$ at $t'=0$ for the NN model of Sec.~\ref{sec:NN_phase}, and $t_A, t'\in\{0.0, 0.1, 0.3, 0.5, 0.7, 0.9\}$ for the combined model. The electron filling $n$ is varied from half filling ($n = 1$) down to quarter-filling ($n = 0.5$). In a square-lattice altermagnet the two magnetic sublattices carry opposite spins, and their surroundings are rotated by $90^\circ$ relative to each other, so the two spin species have different hoppings. In Eqs.~(\ref{eq:H_NN}) and~(\ref{eq:H_NNN}), $t_A$ makes this difference on the $\hat{x}$ and $\hat{y}$ bonds and $t'$ on the two diagonals. The resulting spin splitting has the form $\cos k_x-\cos k_y$ for $t_A$ and $\sin k_x\sin k_y$ for $t'$. A lattice with mirror planes allows only one of these two forms. Without mirror planes both are present, a case that only the combined model describes. 

In a magnetic background the barrier an electron meets along a bond depends on whether the site it moves toward carries the same spin as itself or the opposite one. Because the two sublattices are related by the $90^\circ$ rotation rather than by a translation, the two bond directions present opposite environments to a given spin species, so the effective hopping differs between $\hat{x}$ and $\hat{y}$ and changes sign between them. That difference is $t_A$, and the same argument applied to the two diagonals gives $t'$. Building this pattern needs two magnetic sites in the basis, but the energy scale of the magnetic background is much larger than that of the correlations studied here~\cite{Roig2024}, so the sublattice degree of freedom can be integrated out and its effect kept as a spin-dependent hopping in a single band. The splitting also arises on its own in models where it is not imposed: in a two-orbital model it appears once the interaction that drives orbital order is strong enough~\cite{Leeb2024}, with the same symmetry as the terms used here. In ultracold fermions in optical lattices the anisotropy is engineered directly rather than derived~\cite{Das2024}. Both $t_A$ and $t'$ break time-reversal symmetry while leaving the combined operation of a $90^\circ$ rotation and time reversal intact, so neither is a form of spin-orbit coupling.

The ranges of $t_A$ and $t'$ used here go well beyond the values expected in known compounds, where both hoppings are a small fraction of the bandwidth. No material is expected at $t_A=t'=0.9$. We go up to these values because each effect becomes clear only when its hopping is large enough to dominate. Diagonal correlations cover a large part of the phase diagram only for $t_A\gtrsim0.5$, and stripe correlations only for $t'\gtrsim0.5$. Large values are also meaningful on their own. At large $t_A$ the Fermi surfaces of the two spins become nearly one-dimensional in opposite directions, and at large $t'$ the magnetic frustration is strongest.

Recent studies show that altermagnetic spin splitting can emerge in two-dimensional Hubbard-type systems~\cite{Bai2024, Lee2024, Smejkal2022a, Smejkal2022b}, motivating a systematic many-body investigation of how hopping asymmetry and next-nearest-neighbor frustration jointly shape the magnetic ground state. Eqs.~(\ref{eq:H_NN}) and~(\ref{eq:H_NNN}) are a minimal effective lattice model that isolates the role of bond anisotropy $t_A$ and diagonal hopping $t'$ in reshaping the magnetic phase diagram of a correlated square-lattice system. More elaborate extensions, such as longer-range hopping or spin-orbit coupling, may produce related phenomena but would introduce additional competing mechanisms that obscure the specific role of $t_A$ and $t'$ that we wish to isolate here.

We employ the CPQMC method with a Hartree--Fock trial wave function, which captures quantum fluctuations and many-body correlations in the intermediate-coupling regime~\cite{Mushtaq2026}. By imposing the constrained-path approximation, CPQMC mitigates the fermion sign problem, enabling reliable simulations of ground-state properties across fillings $n \in [0.5, 1.0]$ and parameter combinations $(t_A, t')$. We fix the Hubbard interaction at $U = 4t$ and perform simulations on periodic $L \times L$ clusters, using $L = 14$ for the main results and $L = 8$ and $L = 10$ as finite-size checks (see Appendix~B). We test the constrained-path approximation against exact diagonalization (ED) in Appendix C. The $L = 14$ lattice provides the momentum-space resolution required to track the dominant ordering wavevector across the full $(t_A, t', n)$ parameter space, while the smaller lattices confirm that the magnetic trends and phase boundaries are not dominated by finite-size artifacts.

The trial wave function is the unrestricted Hartree--Fock solution at the same $U$, $t_A$ and $t'$, taken as the lowest of 40 runs started from random densities. Depending on the parameters, it is paramagnetic (close to the free-electron determinant) or spin polarized, as at half filling. Appendix~D compares the results with four other trial states, including the free-electron determinant. The simulations use a time step $\Delta\tau=0.01$ and $1000$ walkers, with population control every $10$--$20$ steps and re-orthogonalization every $10$ steps. Each run has $10$ equilibration and $10$ measurement blocks of $320$ steps. Observables are measured with $40$ steps of back-propagation, and Appendix~C shows how they depend on this length. To check the time-step error, we repeated four runs, one in each magnetic region, at $\Delta\tau=0.005$ and $0.02$ with the same total imaginary time. The energy changed by less than $0.1\%$ and the dominant wavevector $\mathbf{Q}$ did not change, so the phase assignments do not depend on the time step.

\subsection{Magnetic observables}
\label{sec:observables}

To characterize the magnetic ground state, we compute equal-time spin correlations in both momentum and real space. The local spin operator is defined as

\begin{equation}
\hat{S}_i^z = \frac{1}{2}\left(\hat{n}_{i\uparrow} - \hat{n}_{i\downarrow}\right),
\label{eq:Sz_def}
\end{equation}
and the equal-time spin correlation function between sites $i$ and $j$ is

\begin{equation}
C^{zz}(i,j) = \langle \hat{S}_i^z \hat{S}_j^z \rangle.
\label{eq:Czz_def}
\end{equation}

The static spin structure factor is obtained from this as

\begin{equation}
S^z(\mathbf{k}) = \frac{1}{N}\sum_{i,j} e^{i\mathbf{k}\cdot(\mathbf{r}_i - \mathbf{r}_j)}\, C^{zz}(i,j),
\label{eq:Sz_k}
\end{equation}
where $N = L^2$ counts the total number of lattice sites and $\mathbf{r}_i$ ($\mathbf{r}_j$) is the real-space position of site $i$ ($j$). The quantity $S^z(\mathbf{k})$ measures the strength of spin-density-wave (SDW) correlations at wavevector $\mathbf{k}$ and serves as the primary momentum-space diagnostic throughout this work. The dominant magnetic ordering wavevector is identified as

\begin{equation}
\mathbf{Q} = \arg\max_{\mathbf{k}}\, S^z(\mathbf{k}).
\label{eq:Q_def}
\end{equation}

This criterion assigns a unique ordering wavevector to every point in the parameter space. The four resulting magnetic correlations and their associated $\mathbf{Q}$ vectors are summarized in Table~\ref{tab:classification}. The names in Table~\ref{tab:classification} refer to the wavevector of the $S^z(\mathbf{k})$ peak only. In particular, spiral labels a peak at $(\pi,q)$ or $(q,\pi)$ with $q\neq0,\pi$, following Ref.~\cite{Mushtaq2026}. $S^z(\mathbf{k})$ alone does not determine the spin arrangement behind such a peak, and no statement about it is made. We use the terms phase diagram and phase boundary for the map of the dominant correlation and for the lines where the dominant wavevector changes. Away from half filling these regions are correlation regimes, not ordered phases (Appendix~B).

\begin{table}[htb]
\centering
\caption{Magnetic correlation classification by dominant ordering wavevector $\mathbf{Q}$, extracted from the global maximum of $S^z(\mathbf{k})$.}
\begin{tabular}{lc}
\hline\hline
Magnetic correlation & Ordering wavevector $\mathbf{Q}$ \\
\hline
N\'{e}el antiferromagnetism   & $(\pi,\pi)$ \\
Stripe magnetism              & $(\pi,0)$ or $(0,\pi)$ \\
Diagonal spin-density-wave    & $(q,q)$,\quad $q\neq\pi$ \\
Spiral magnetism              & $(\pi,q)$ or $(q,\pi)$,\quad $q\neq 0,\pi$ \\
\hline\hline
\end{tabular}
\label{tab:classification}
\end{table}

Distinguishing genuine long-range magnetic order from short-range correlations requires combining momentum-space and real-space information. In momentum space, a phase with true long-range order displays a narrow, well-resolved peak in $S^z(\mathbf{k})$ at $\mathbf{Q}$ whose integrated weight grows with system size, while short-range correlations produce a broad spectral feature near $\mathbf{Q}$ whose width scales inversely with the magnetic correlation length. To quantify this sharpness, we define the sharpness index

\begin{figure*}[!t]
    \centering
    \includegraphics[width=0.8\textwidth]{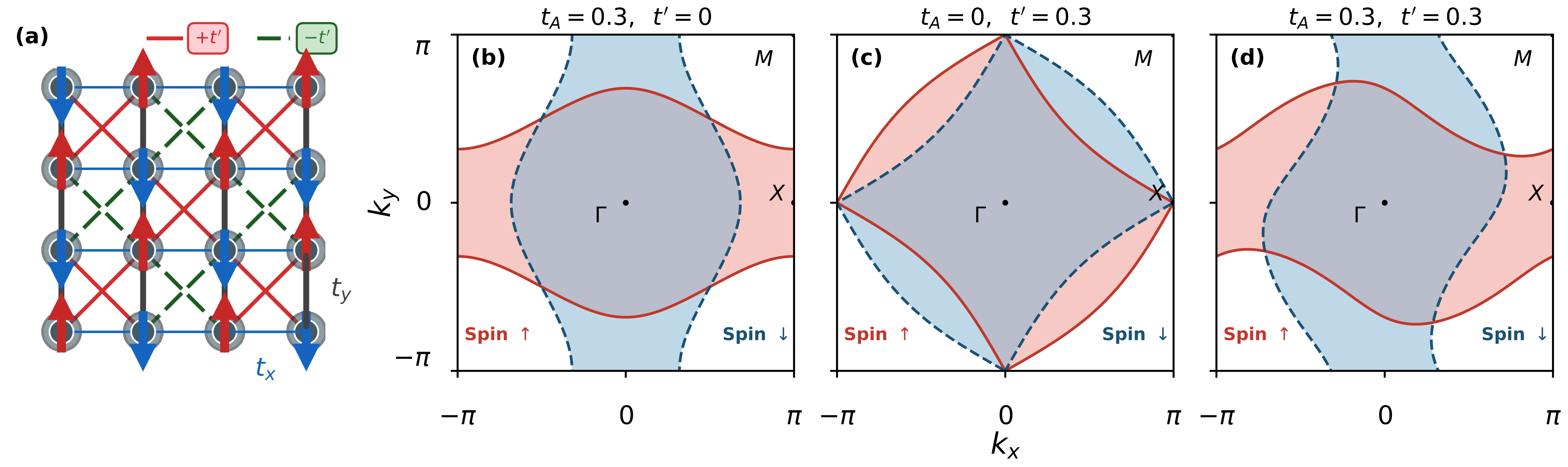}
    \caption{(a) Schematic of the altermagnetic square-lattice model with spin-dependent NN hopping anisotropy $t_A$ and NNN hopping $t^{\prime}$. The hopping is opposite for the two spin species along $\hat{x}$ and $\hat{y}$, inducing a $d$-wave momentum-dependent spin splitting at the single-particle level. (b)--(d) Spin-resolved Fermi surfaces at half filling for three representative 
    parameter sets: (b) $t_A = 0.3,\ t^{\prime} = 0$; (c) $t_A = 0,\ t^{\prime} = 0.3$; (d) $t_A = 0.3,\ t^{\prime} = 0.3$. Red (blue) shading and lines denote spin-up (spin-down) occupied regions and Fermi contours, respectively.}
    \label{fig:model}
\end{figure*}

\begin{equation}
R_p = \frac{S^z(\mathbf{Q}) - \bar{S}^z_{\mathrm{near}}(\mathbf{Q})}{S^z(\mathbf{Q})},
\label{eq:Rp_def}
\end{equation}
where $\bar{S}^z_{\mathrm{near}}(\mathbf{Q})$ is the average of $S^z(\mathbf{k})$ over the NN $\mathbf{k}$-points surrounding $\mathbf{Q}$. An $R_p$ value near unity signals a sharp, well-developed peak consistent with strong magnetic ordering, while $R_p \approx 0$ indicates a broad or featureless response. Because $R_p$ compares the peak with its neighbors on the momentum grid, it measures how sharp the peak is at a fixed $L$. It is not an order parameter, so we judge long-range order only from the size dependence of $S^{z}(\mathbf{Q})/N$ in Appendix~B. The real-space information is extracted from the translationally averaged SDW correlation function

\begin{equation}
C_{\mathrm{SDW}}(\mathbf{r}) = \frac{1}{N}\sum_j \langle \hat{S}_{j+\mathbf{r}}^z \hat{S}_j^z \rangle,
\label{eq:CSDW_def}
\end{equation}
where $\mathbf{r}$ is the displacement between two sites. In real space, long-range order is signaled by sustained oscillations in $C_{\rm SDW}(\mathbf{r})$ that retain a finite amplitude across the full lattice, whereas short-range correlations fall off exponentially at large separations, indicating the absence of macroscopic phase coherence. By combining the sharpness of the $S^z(\mathbf{k})$ peak through $R_p$ with the asymptotic behavior of $C_\mathrm{SDW}(\mathbf{r})$, the nature and extent of magnetic correlations can be reliably identified across the full $(n, t_A, t')$ parameter space.

\section{Noninteracting Electronic Structure}
\label{sec:noninteracting}

Before analyzing the interacting magnetic correlations, it is useful to first examine how $t_A$ and $t^{\prime}$ reshape the noninteracting electronic structure. Because the two spin species experience opposite hopping anisotropy along $\hat{x}$ and $\hat{y}$, the band structure develops a momentum-dependent spin splitting, the spin-resolved Fermi surfaces deform, and the associated van Hove energies move apart~\cite{VanHove1953, YuanFu2020}. The NNN hopping $t^{\prime}$ further reshapes the spin-resolved bands through the $\sin k_x \sin k_y$ term. When both parameters are nonzero, the two effects combine and reshape the low-energy phase space more than either term does alone. The analysis below should therefore be viewed as the interacting consequence of a band structure already reorganized at the single-particle level.

\subsection{Momentum-dependent spin splitting and band reconstruction}

The noninteracting part of $\hat{H}_\text{NN} + \hat{H}_\text{NNN}$ is diagonalized in momentum space, yielding spin-resolved single-particle dispersions
\begin{align}
    \varepsilon_{\uparrow}(\mathbf{k}) &= -2\left[(t-t_A)\cos k_x 
    + (t+t_A)\cos k_y\right] \nonumber\\
    &\quad + 4t^{\prime}\sin k_x \sin k_y, \label{eq:disp_up}\\
    \varepsilon_{\downarrow}(\mathbf{k}) &= -2\left[(t+t_A)\cos k_x 
    + (t-t_A)\cos k_y\right] \nonumber\\
    &\quad - 4t^{\prime}\sin k_x \sin k_y. \label{eq:disp_dn}
\end{align}
where $t = 1$ is the energy unit, $t_A$ controls the spin-dependent NN hopping anisotropy along $\hat{x}$ and $\hat{y}$, and $t^{\prime}$ is the spin-dependent NNN hopping. In the limit $t_A = t^{\prime} = 0$, Eqs.~(\ref{eq:disp_up})--(\ref{eq:disp_dn}) reduce to the conventional square-lattice dispersion $\varepsilon(\mathbf{k}) = -2t(\cos k_x + \cos k_y)$, which is spin-degenerate and exhibits perfect particle-hole symmetry and $(\pi,\pi)$ Fermi-surface nesting at half filling. Finite $t_A$ or $t^{\prime}$ lifts this degeneracy in qualitatively distinct ways, as described below.

The spin-resolved Fermi surfaces at half filling are shown in Fig.~\ref{fig:model}(b)--(d) for three representative parameter sets. When only NN anisotropy is present, the spin-up and spin-down Fermi sheets retain the connected topology of the isotropic limit but deform anisotropically in opposite directions. The spin-up sheet elongates along $k_x$ while the spin-down sheet elongates along $k_y$. Perfect $(\pi,\pi)$ nesting is preserved, as $t_A$ alone does not break particle-hole symmetry. When only NNN hopping is present, the spin splitting is driven entirely by the $4t^{\prime}\sin k_x\sin k_y$ term, which vanishes along the zone axes ($k_x = 0$ or $k_y = 0$) and the zone boundary ($k_x = \pm\pi$ or $k_y = \pm\pi$). The Fermi-surface deformation is therefore concentrated near the $(\pi,0)$ and $(0,\pi)$ saddle points, where the spin-up and spin-down sheets deform in opposite directions in the Brillouin zone interior. In the combined case, both mechanisms reinforce each other, producing spin-selective anisotropy across the entire Brillouin zone. This cumulative deformation of the Fermi surface modifies the low-energy particle-hole scattering channels and provides the single-particle foundation for the magnetic correlation evolution.

\subsection{Van Hove singularities}
\label{VHS_distribution}

\begin{figure*}[!t]
  \centering
  \includegraphics[width=0.8\textwidth]{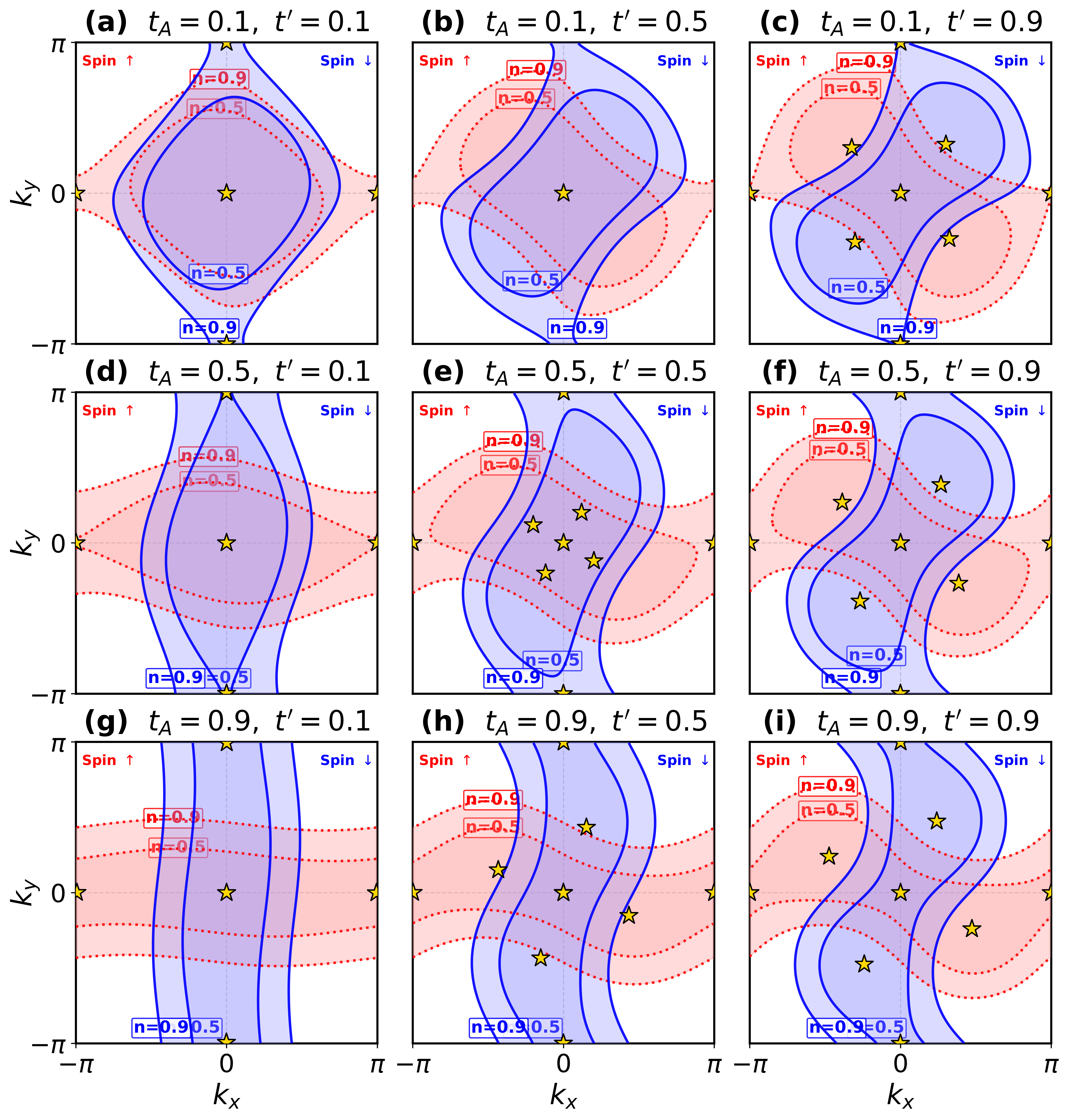}
  \caption{Evolution of the Fermi surface and van Hove singularity locations across the $(t_A,\,t^{\prime})$ parameter space. Rows correspond to $t_A=0.1,\,0.5,\,0.9$ (top to bottom) and columns to $t^{\prime}=0.1,\,0.5,\,0.9$ (left to right). In each panel, red dotted (blue solid) contours show the spin-up (spin-down) Fermi surface at filling $n=0.5$ and $n=0.9$, with occupied regions shaded. Gold stars mark the van Hove points $\nabla_{\mathbf{k}}\varepsilon(\mathbf{k})=0$. Once $t_A^2+4t'^2>1$, the band at $\Gamma$ and $M$ turns from a minimum or maximum into a saddle point, and new minima and maxima (stars) appear inside the zone; the saddle points at $X$ stay. Increasing $t_A$ progressively deforms the two spin Fermi sheets anisotropically in opposite directions.}
    \label{fig:vhs_fs}
\end{figure*}

\begin{figure*}[!t]
    \centering
    \includegraphics[width=0.74\textwidth]{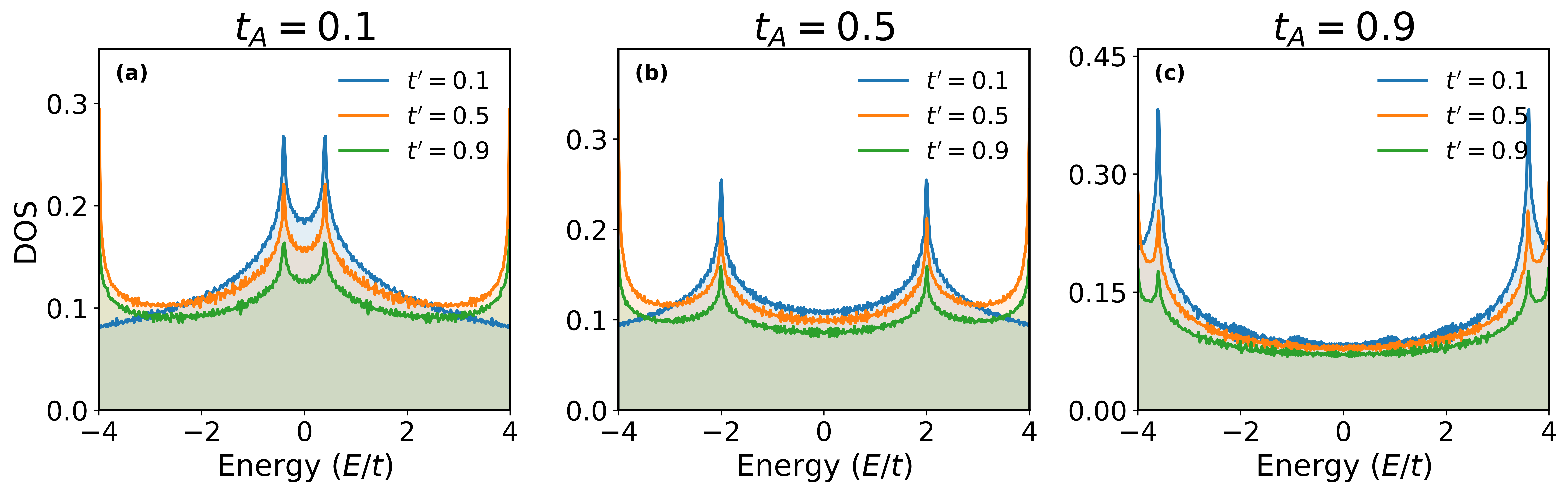}
    \caption{Density of states as a function of energy $E/t$ for (a)~$t_A=0.1$, (b)~$t_A=0.5$, and (c)~$t_A=0.9$. In each panel, curves correspond to $t^{\prime}=0.1$ (blue), $0.5$ (orange), and $0.9$ (green). At small $t^{\prime}$, the DOS shows a conventional logarithmic van Hove peak near the band center. At large $t^{\prime}$, the single peak splits into multiple features. Increasing $t_A$ (left to right) progressively shifts spectral weight away from $E=0$, consistent with the spin-split van Hove energies $\pm 4\,t_A$.}
    \label{fig:dos}
\end{figure*}

Having established the symmetry structure of the spin-resolved dispersions, we next examine how $t_A$ and $t^{\prime}$ jointly reshape the noninteracting Fermi surfaces and van Hove singularities (VHS), and whether their effects add independently or couple as either is tuned. The spin-resolved Fermi contours across the $(t_A, t^{\prime})$ parameter space are shown in Fig.~\ref{fig:vhs_fs} for two representative fillings, $n=0.5$ and $n=0.9$, where red dotted (blue solid) lines denote spin-up (spin-down). Gold stars mark the van Hove points, identified as the stationary points $\nabla_{\mathbf{k}}\varepsilon(\mathbf{k})=0$~\cite{VanHove1953}.

Moving along the columns of Fig.~\ref{fig:vhs_fs}, i.e., increasing $t^{\prime}$ at fixed $t_A$, one observes a qualitative change in the Fermi surface shape and in the van Hove points of the band. At small $t^{\prime}$, the van Hove singularities sit at the zone-boundary $X$ points $(\pi,0)$ and $(0,\pi)$, and the Fermi contours retain the connected topology of the isotropic limit. As $t^{\prime}$ increases, new van Hove points (stars) appear inside the Brillouin zone, while the saddle points at $X$ stay. The point where this change happens can be found analytically from the band curvature at the zone corner $M=(\pi,\pi)$. There, the dispersion $\varepsilon_\sigma(\mathbf{k})$ turns from a band extremum into a saddle point exactly when $t_A^2+4t'^2=1$, which reduces to $t'=1/2$ at $t_A=0$. At $(\pi,\pi)$ the band has zero slope for all $t_A$ and $t'$, and the determinant of its second derivatives is $-4(t_A^2+4t'^2-1)$. The point is therefore a band minimum or maximum for $t_A^2+4t'^2<1$ and a saddle point beyond. The same holds at $\Gamma$, and the new minima and maxima of Fig.~\ref{fig:vhs_fs} split off from $\Gamma$ and $M$ there. This is a Lifshitz-type change of the band topology~\cite{Lifshitz1960}. Because $\Gamma$ and $M$ lie at the band edges, $\varepsilon=\mp4t$, this change does not affect the Fermi surface for $0.5\le n\le1$ and plays no role below. The Fermi level does cross the van Hove energies $\mp4t_A$ of the $X$ points inside the studied range: at $n\approx0.84$--$0.89$ for $t_A=0.1$, at lower filling for larger $t_A$, and outside the range for $t_A\gtrsim0.6$. These crossings do not coincide with the phase boundaries of Sec.~\ref{magnetic_correlations}. In the combined case, $t_A$ deforms the Fermi surfaces while $t^{\prime}$ reshapes the band, and together they produce spin-selective reconstruction across the Brillouin zone.

Moving along the rows of Fig.~\ref{fig:vhs_fs}, i.e., increasing $t_A$ at fixed $t^{\prime}$, produces a qualitatively different effect. $t_A$ does not move the van Hove momenta from the $X$ points, but it shifts the VHS energies of the two spin channels in opposite directions such as $\varepsilon_{\uparrow}(\pi,0)=-4\,t_A$ and $\varepsilon_{\downarrow}(\pi,0)=+4\,t_A$, with a total spin splitting of $8\,t_A$. The primary effect visible in the figure is therefore an anisotropic deformation of the Fermi sheets. One spin species is progressively stretched along $\hat{k}_y$ while the other is stretched along $\hat{k}_x$, and this deformation intensifies with increasing $t_A$. At large $t_A$, the two spin sectors acquire strongly opposite orientations in momentum space, consistent with the quasi-one-dimensional character discussed in the context of Fig.~\ref{fig:model}. The reconstruction also depends on filling. At $n=0.9$, the Fermi contour approaches the zone boundary and the deformation sets in earlier and more strongly than at $n=0.5$, where the contour remains further from the saddle points. Finite $t_A$ therefore does not affect all densities in the same way; the extent of the reconstruction depends on how the Fermi level samples the anisotropic dispersion.

When both $t_A$ and $t^{\prime}$ are nonzero, the two mechanisms act cooperatively. The $t^{\prime}$ term reshapes the band, while $t_A$ simultaneously separates the energy scales of the two spin channels. In the NNN-only model all of the spin splitting comes from $t^{\prime}$, and the NN bands of the two spins are identical. Here $t_A$ splits the two spins at any $t^{\prime}$. The spin splitting is therefore set by $t_A$ and $t^{\prime}$ together across the full $(n, t_A, t^{\prime})$ range, not by $t^{\prime}$ alone as in the NNN-only model.

The same reconstruction is reflected in the total density of states (DOS) shown in Fig.~\ref{fig:dos}. At small $t^{\prime}$ (blue curves), the DOS exhibits a conventional logarithmic peak near the band center, with the van Hove singularities remaining at the $X$ points. As $t^{\prime}$ increases toward the intermediate value (orange curves), the peak sharpens and narrows. At large $t^{\prime}$, the single peak splits into multiple features. The spectral weight near $E=0$ is progressively suppressed and redistributed toward the band edges as $t_A$ increases, reflecting the spin-split VHS energies $\pm 4\,t_A$. Taken together, the Fermi-surface evolution in Fig.~\ref{fig:vhs_fs} and the DOS evolution in Fig.~\ref{fig:dos} provide a complete noninteracting picture of how the combined altermagnetic parameters reorganize the low-energy phase space.

\section{Evolution of Magnetic Correlations}
\label{magnetic_correlations}

To characterize the magnetic correlations in the interacting system, we compute the equal-time spin structure factor $S^z(\mathbf{k})$ defined in Eq.~(\ref{eq:Sz_k}). Magnetic correlations are identified from the dominant peak structure of $S^z(\mathbf{k})$ using a systematic classification. For each parameter point $(n, t_A, t')$, we identify the dominant magnetic correlation from the position of the global maximum of $S^z(\mathbf{k})$, classifying it as N\'{e}el at $\mathbf{Q} = (\pi,\pi)$, stripe at $\mathbf{Q} = (\pi,0)$ or $(0,\pi)$, spiral at $\mathbf{Q} = (\pi,q)$ or $(q,\pi)$, or diagonal at $\mathbf{Q} = (q,q)$. For both spiral and diagonal points, the wavevector component $q$ is extracted directly as the peak coordinate, along the zone boundary for spiral correlations and along the zone diagonal for diagonal correlations. This $q$ value is shown explicitly in the phase diagrams as a continuous color scale. The analysis starts with the most basic scenario, which is the pure NN altermagnetic model with $t' = 0$, and moves on to the full combined model before reducing back to the NNN limit to connect with earlier research.

\subsection{Phase diagram of the NN altermagnetic model}
\label{sec:NN_phase}

\begin{figure*}[!t]
    \centering
    \includegraphics[width=0.76\textwidth]{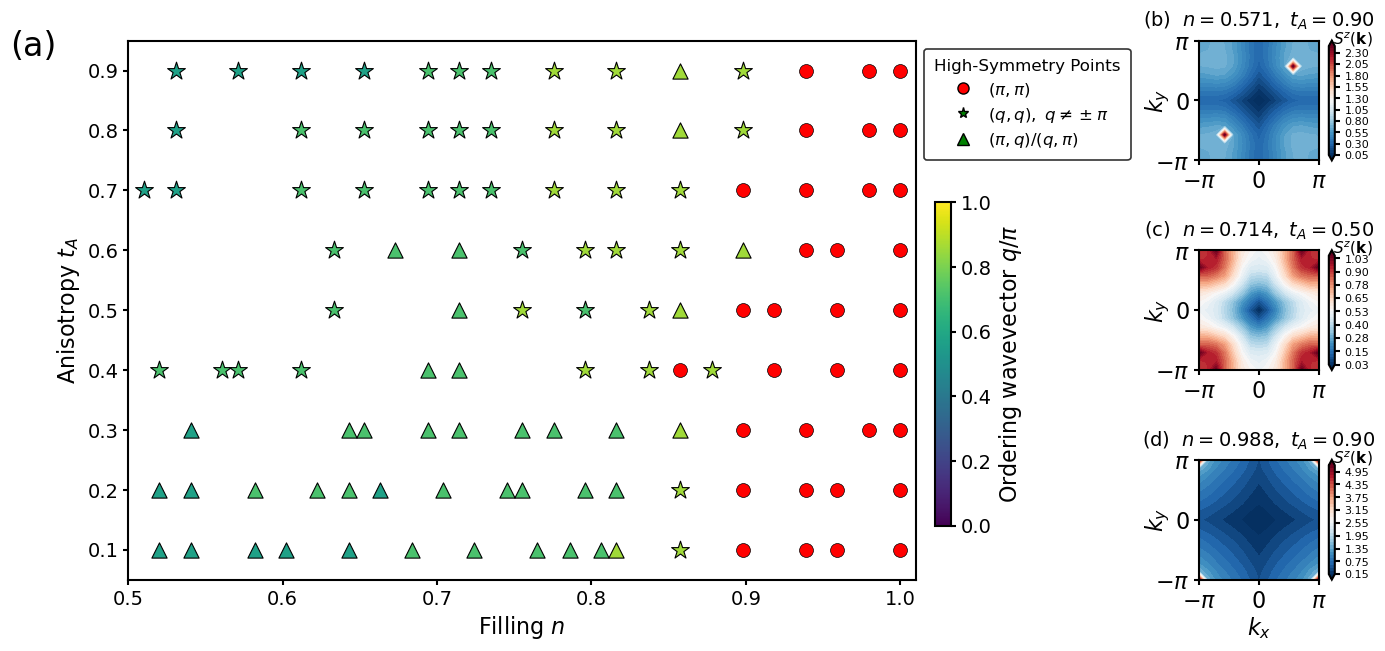}
    \caption{Magnetic phase diagram and representative $S^z(\mathbf{k})$ maps of the NN altermagnetic Hubbard model ($U=4t$, $L=14$, $t^{\prime}=0$). Dominant $S^z(\mathbf{k})$ peak positions in the $(n,\,t_A)$ plane are shown in panel (a), classifying each point as N\'{e}el correlations $\mathbf{Q}=(\pi,\pi)$, spiral correlations $\mathbf{Q}=(\pi,q)$ or $(q,\pi)$, or diagonal correlations $\mathbf{Q}=(q,q)$. The colorbar encodes the ordering wavevector $q/\pi$. Near half filling, N\'{e}el correlations persist for all $t_A$. Upon hole doping, the system transitions to spiral correlations at weak anisotropy and to diagonal correlations at strong anisotropy. Panels (b), (c), and (d) show representative $S^z(\mathbf{k})$ maps for the diagonal ($n=0.571$, $t_A=0.90$), spiral ($n=0.714$, $t_A=0.50$), and N\'{e}el ($n=0.988$, $t_A=0.90$) correlations, respectively.}
    \label{fig:phase_NN}
\end{figure*}

We begin with the CPQMC results for the pure NN altermagnetic model described by $\hat{H}_\text{NN}$ in Eq.~\eqref{eq:H_NN}, at $L = 14$, $U=4t$ and $t^{\prime}=0$. The dominant $S^z(\mathbf{k})$ peak locations in the $(n,\,t_A)$ plane are shown in Fig.~\ref{fig:phase_NN}(a), and representative $S^z(\mathbf{k})$ maps for the three identified correlations are shown in Figs.~\ref{fig:phase_NN}(b)--(d). The figure reveals a clear evolution among three distinct magnetic correlations, namely N\'{e}el correlations at $\mathbf{Q}=(\pi,\pi)$, spiral correlations at $\mathbf{Q}=(\pi,q)$ or $(q,\pi)$, and diagonal correlations at $\mathbf{Q}=(q,q)$ with $q\neq\pm\pi$, as the filling $n$ and hopping anisotropy $t_A$ are varied. At half filling ($n=1$) the dominant correlation is N\'{e}el, $\mathbf{Q}=(\pi,\pi)$, for all values of $t_A$ studied, with sharp $S^z(\mathbf{k})$ peaks at the zone corners. Finite $t_A$ shifts the VHS energies of the two spin channels to $\pm 4\,t_A$ but leaves the van Hove momenta pinned at the $X$ points $(\pi,0)$ and $(0,\pi)$, so the perfect $(\pi,\pi)$ Fermi-surface nesting is preserved (see Appendix~A for a detailed VHS analysis of the pure NN case).

Away from half filling, the dominant ordering vector evolves in a manner that depends on $t_A$. At weak anisotropy ($t_A \lesssim 0.3$), doping suppresses the N\'{e}el peak and the system develops spiral correlations at $\mathbf{Q}=(\pi,q)$ or $(q,\pi)$. The corresponding $S^z(\mathbf{k})$ map in Fig.~\ref{fig:phase_NN}(c) shows spectral weight distributed along the zone boundary with a peak displaced from the N\'{e}el point, confirming the spiral character. The colorbar encodes $ q/\pi$, with triangles closer to yellow indicating $q$ approaching $\pi$, while greener triangles indicate a more pronounced departure from the N\'{e}el point. This evolution is driven by the gradual loss of $(\pi,\pi)$ Fermi-surface nesting with doping, which shifts the dominant particle-hole susceptibility toward spiral wave vectors. At larger anisotropy ($t_A \gtrsim 0.5$), the dominant ordering tendency shifts toward diagonal correlations $\mathbf{Q}=(q,q)$, visible as the stars in Fig.~\ref{fig:phase_NN}(a), which are colored by the same $q/\pi$ scale as the spiral triangles. The $S^z(\mathbf{k})$ map in Fig.~\ref{fig:phase_NN}(b) clearly shows the spectral weight concentrated along the diagonal direction, distinct from both the N\'{e}el and spiral correlations. This diagonal character arises because the strongly deformed spin-resolved Fermi sheets at large $t_A$ acquire opposite elongations along $\hat{k}_x$ and $\hat{k}_y$, redirecting the dominant particle-hole scattering away from the $(\pi,\pi)$ channel toward the diagonal sector.

A second important feature of the phase diagram is the filling range over which N\'{e}el correlations persist. At small $t_A$, the N\'{e}el phase extends down to relatively low fillings before being replaced by spiral correlations, whereas at large $t_A$ the N\'{e}el region narrows and the transition sets in already at higher filling. Physically, increasing $t_A$ separates the van Hove energies of the two spin channels by $8\,t_A$, progressively redistributing spectral weight away from the band center and weakening the particle-hole phase space that stabilizes N\'{e}el correlations. Taken together, the $(n,\,t_A)$ phase diagram and the representative $S^z(\mathbf{k})$ maps show that NN anisotropy alone drives competition among three correlations. We turn to the combined model next.

\subsection{Phase diagram of the combined model}
\label{sec:combined_phase}

\begin{figure*}[!t]
    \centering
    \includegraphics[width=0.8\textwidth]{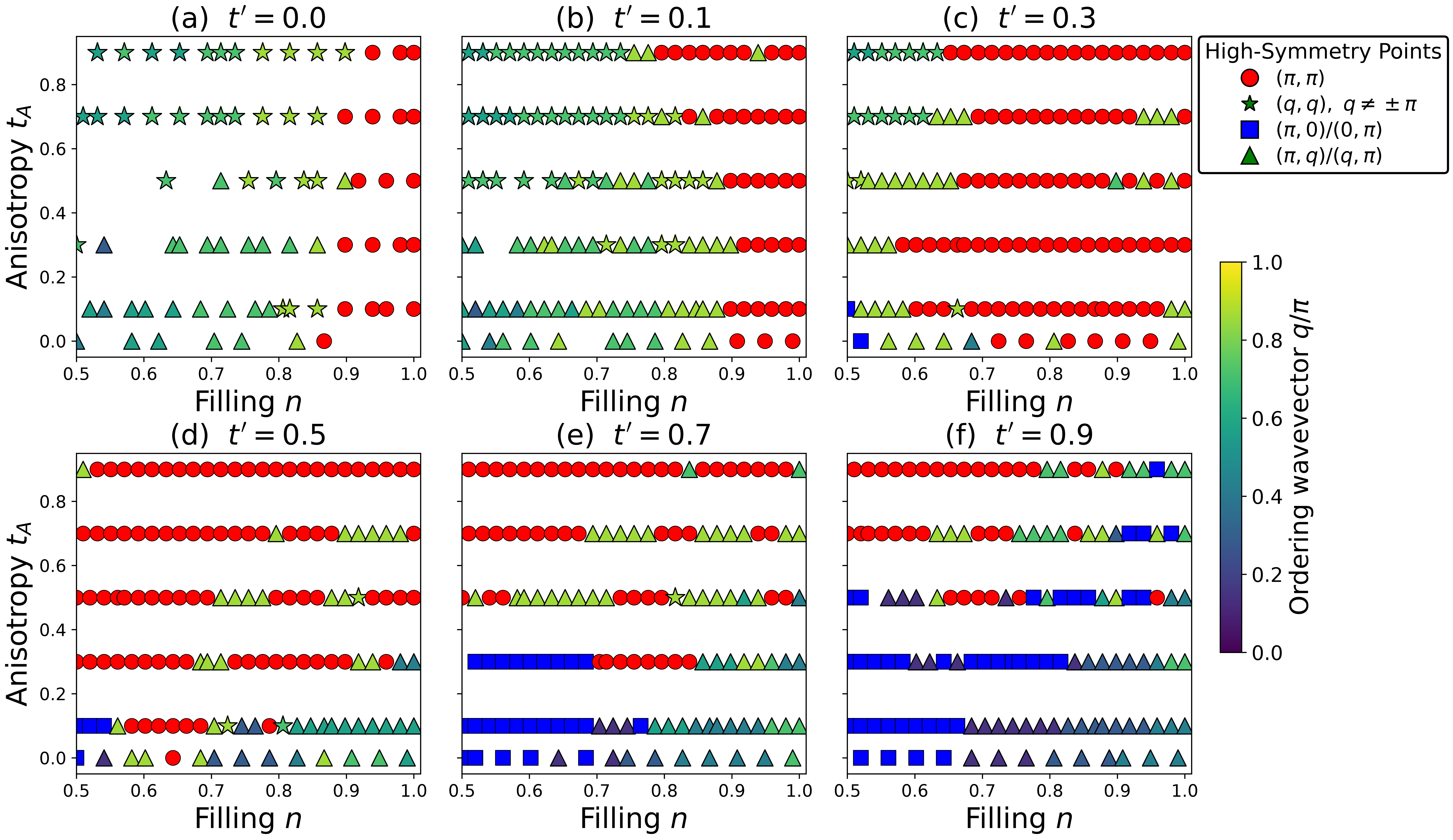}
    \caption{Magnetic phase diagram of the combined altermagnetic Hubbard model ($U=4t$, $L=14$) in the $(n,\,t_A)$ plane. Panels~(a)--(f) correspond to $t^{\prime}=0.0$, $0.1$, $0.3$, $0.5$, $0.7$, and $0.9$. The dominant $S^z(\mathbf{k})$ peak position classifies each point as N\'{e}el correlations $\mathbf{Q}=(\pi,\pi)$, spiral correlations $\mathbf{Q}=(\pi,q)/(q,\pi)$, stripe correlations $\mathbf{Q}=(\pi,0)/(0,\pi)$, or diagonal correlations $\mathbf{Q}=(q,q)$ with $q\neq\pm\pi$; the colorbar encodes the ordering wavevector $q/\pi$. Panel~(a) at $t^{\prime}=0$ shows the NN altermagnetic model of Fig.~\ref{fig:phase_NN} at the six values of $t_A$ used for the combined model and serves as the baseline. As $t^{\prime}$ increases, stripe correlations emerge at small $t_A$ and spread across the phase diagram, while increasing $t_A$ reinforces N\'{e}el correlations and stabilizes the diagonal correlations.}
    \label{fig:phase_combined}
\end{figure*}

We now extend the analysis to the full combined model in which both $t_A$ and $t^{\prime}$ are varied. The dominant $S^z(\mathbf{k})$ peak locations in the $(n,\,t_A)$ plane are shown in Fig.~\ref{fig:phase_combined}, where panels~(a)--(f) correspond to increasing NNN hopping $t^{\prime}=0.0$, $0.1$, $0.3$, $0.5$, $0.7$, and $0.9$. Panel~(a), at $t^{\prime}=0$, shows the NN altermagnetic model of Fig.~\ref{fig:phase_NN} at the six values of $t_A$ of the combined model and provides the baseline against which the effect of finite $t^{\prime}$ can be assessed. Four correlations appear across the full data set. N\'{e}el correlations at $\mathbf{Q}=(\pi,\pi)$ occupy the high-filling region; spiral correlations at $\mathbf{Q}=(\pi,q)$ or $(q,\pi)$ develop at intermediate filling; stripe correlations at $\mathbf{Q}=(\pi,0)$ or $(0,\pi)$ appear at small $t_A$ and large $t^{\prime}$; and diagonal correlations at $\mathbf{Q}=(q,q)$ appear at large $t_A$ and small $t^{\prime}$.

As $t^{\prime}$ is increased from panel~(a) to panel~(f), two systematic trends emerge. First, stripe correlations, which are absent at $t^{\prime}=0$, appear at small $t_A$ once $t^{\prime}$ is turned on and spread to occupy a growing low-$t_A$ region of the phase diagram. By $t^{\prime}=0.9$, the blue-square stripe region covers the small-$t_A$ rows across nearly the full filling range. Stripe correlations are therefore driven entirely by $t^{\prime}$, emerging only when the NNN hopping is strong enough to redirect the dominant particle-hole scattering toward the $(\pi,0)/(0,\pi)$ channel. Second, the diagonal correlations, marked by stars at large $t_A$ and small $t^{\prime}$ in panels~(a)--(c), are progressively suppressed as $t^{\prime}$ grows and have largely given way to N\'{e}el and spiral correlations by $t^{\prime}=0.5$. The diagonal correlations thus require both finite $t_A$ and small $t^{\prime}$ to survive.

The evolution of the N\'{e}el phase ties these trends together. At $t^{\prime}=0$, N\'{e}el correlations occupy the high-filling region for all $t_A$. As $t^{\prime}$ increases, the N\'{e}el region splits along the $t_A$ axis: at large $t_A$ it survives and extends to lower filling, while at small $t_A$ it is replaced by stripe correlations. Reading each panel from bottom to top at fixed large $t^{\prime}$, the dominant correlation therefore evolves from stripe at small $t_A$, through spiral, to N\'{e}el at large $t_A$, showing that $t_A$ reinforces the $(\pi,\pi)$ tendency while $t^{\prime}$ works against it. Taken together, the six panels show that $t_A$ and $t^{\prime}$ act cooperatively to reshape the magnetic phase diagram. $t^{\prime}$ drives the system from N\'{e}el toward spiral and stripe correlations along the frustration axis, while $t_A$ reinforces N\'{e}el correlations, suppresses stripe correlations, and stabilizes the diagonal correlations. The qualitative phase boundaries are consistent with the $L=8$ and $L=10$ results presented in Appendix~B.

The statistical error of the dominant $S^z(\mathbf{k})$ peak is small, with a median of $0.3\%$ of the peak height on the $L=14$ lattice. At most parameter points this peak is higher than the largest peak of any other correlation type by more than two error bars. The remaining points lie on the boundaries in Fig.~\ref{fig:phase_combined}, mostly between the spiral and diagonal regions at small $t'$. There the two peaks are nearly equal, and the boundaries are uncertain by one grid point in $n$.

\begin{figure*}[!t]
    \centering
    \includegraphics[width=0.6\textwidth]{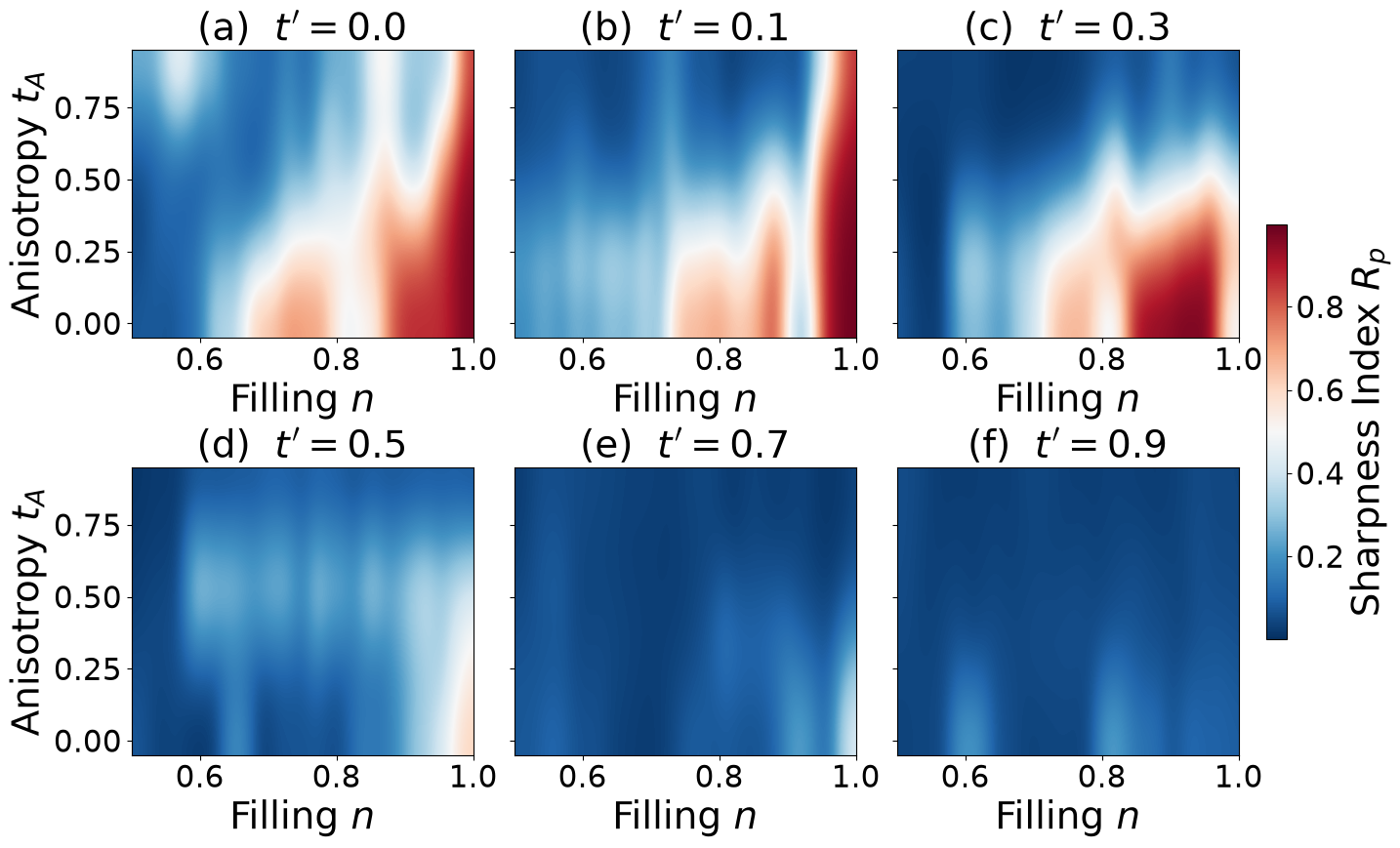}
    \caption{Peak sharpness index $R_p$ of the combined altermagnetic Hubbard model ($U=4t$, $L=14$) in the $(n,\,t_A)$ plane. Panels~(a)--(f) correspond to $t^{\prime}=0.0$, $0.1$, $0.3$, $0.5$, $0.7$, and $0.9$. Red (blue) indicates a sharp (broad) $S^z(\mathbf{k})$ peak. The sharpest response appears near half filling at small $t^{\prime}$, and $R_p$ decreases as $t^{\prime}$ grows or as the system is doped away from half filling.}
    \label{fig:sharpness_combined}
\end{figure*}

To quantify how well-defined the magnetic ordering wavevector is at each parameter point, we define the peak sharpness index $R_p$. For each $(n, t_A, t')$, we locate the wavevector $\mathbf{Q}_\mathrm{max}$ where the spin structure factor $S^z(\mathbf{k})$ is maximum with value $S^z_\mathrm{max}$. The sharpness index ranges from 0 to 1. A value near 1 means the neighboring $\mathbf{k}$-points carry negligible weight, indicating sharp, well-defined magnetic correlations at a single wavevector. A value near 0 means the neighboring $\mathbf{k}$-points carry nearly as much weight as the peak, indicating a broad or diffuse spin response consistent with weak or short-ranged correlations.

Figure~\ref{fig:sharpness_combined} shows $R_p$ across the $(n, t_A)$ plane for the same six values of $t^{\prime}$. The highest sharpness values appear near half filling at small $t^{\prime}$ in panels~(a)--(c), where the spin structure factor develops a sharp and well-defined peak. Reducing $n$ or increasing $t^{\prime}$ lowers $R_p$, indicating a broader and more diffuse spin response, and by $t^{\prime}=0.7$ the response is broad across the entire plane. At fixed $t^{\prime}$, the sharp region tracks the N\'{e}el correlations, confirming that the most well-defined magnetic order develops where the $(\pi,\pi)$ channel dominates.

The steps in some phase boundaries come from the finite momentum grid, not from the constrained-path approximation. On an $L\times L$ lattice the allowed wavevectors are spaced by $2\pi/L$, so a wavevector that changes smoothly with filling or hopping can only appear at the nearest allowed value. The jumps in $q$ between neighboring fillings are exactly $2\pi/L$ on all three lattices. The N\'eel and stripe wavevectors exist on every grid, so their boundaries are not affected.

\subsection{Spatial range of the magnetic correlations}
\label{sec:Rp_real}

\begin{figure*}[!t]
    \centering
    \includegraphics[width=0.79\textwidth]{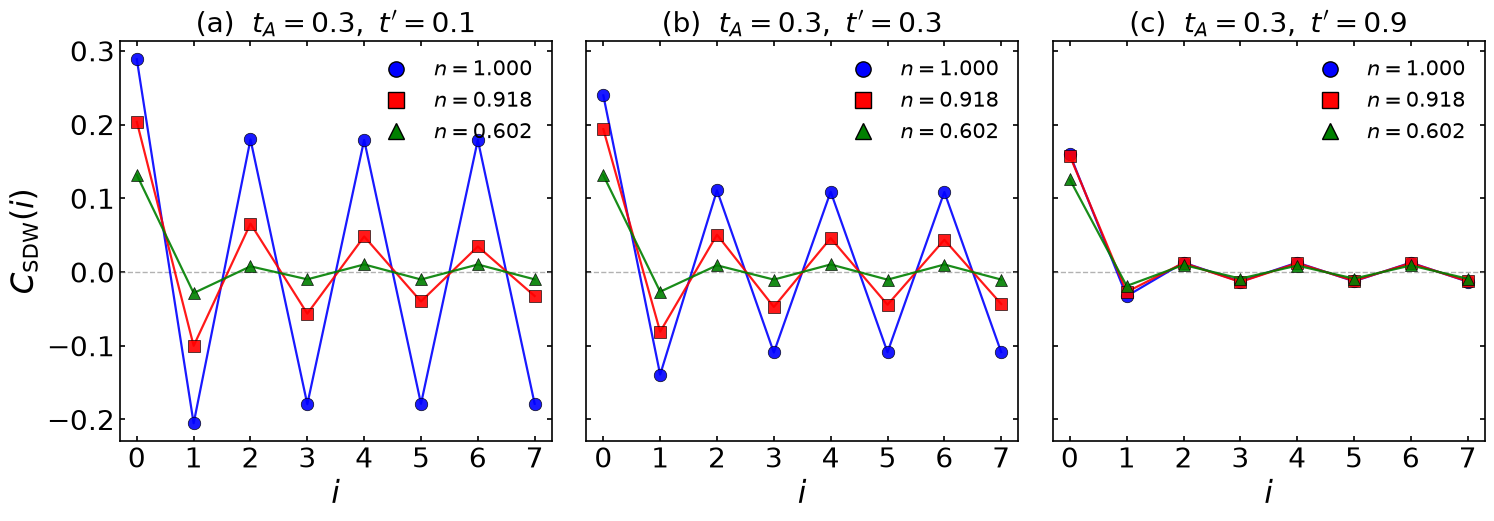}
    \caption{Real-space SDW correlator $C_{\mathrm{SDW}}(i)$ along the $x$ direction at fixed $t_A=0.3$ for three fillings, $n=1.000$, $n=0.918$, and $n=0.602$, at NNN hopping (a)~$t^{\prime}=0.1$, (b)~$t^{\prime}=0.3$, and (c)~$t^{\prime}=0.9$. At small $t^{\prime}$, the correlations near half filling sustain sign-alternating oscillations across the lattice, whereas those at low filling decay within one or two sites. Increasing $t^{\prime}$ suppresses the oscillations at all fillings, and by $t^{\prime}=0.9$ even the half filling correlations are short-ranged, signaling the loss of long-range order.}
    \label{fig:csdw_rp}
\end{figure*}

The sharpness index $R_p$ defined in Eq.~(\ref{eq:Rp_def}) and mapped in Fig.~\ref{fig:sharpness_combined} quantifies how concentrated the spin structure factor is at its dominant wavevector. To confirm that it correctly reflects the spatial range of the magnetic correlations, we compare it directly with the real-space correlator $C_{\mathrm{SDW}}(i)$ of Eq.~(\ref{eq:CSDW_def}), evaluated along the $x$ direction. Figure~\ref{fig:csdw_rp} shows $C_{\mathrm{SDW}}(i)$ for three representative fillings, $n=1.000$, $0.918$, and $0.602$, at fixed $t_A=0.3$ and increasing NNN hopping $t^{\prime}=0.1$, $0.3$, and $0.9$.

At small $t^{\prime}$ in Fig.~\ref{fig:csdw_rp}(a), the three fillings behave very differently. At $n=1.000$, where $R_p$ is largest, $C_{\mathrm{SDW}}(i)$ exhibits sign-alternating oscillations that retain a finite amplitude across the full lattice, the real-space signature of long-range N\'{e}el order~\cite{Xu2022}. At $n=0.918$, where $R_p$ is intermediate, the oscillations are weaker but still persist over several lattice spacings, while at $n=0.602$, where $R_p$ is small, $C_{\mathrm{SDW}}(i)$ decays to zero within one or two sites. A larger $R_p$ therefore goes together with correlations that reach further across the lattice, showing that $R_p$ is a reliable measure of how far the magnetic order extends.

Increasing $t^{\prime}$ at fixed filling progressively suppresses these oscillations. At $t^{\prime}=0.3$ in Fig.~\ref{fig:csdw_rp}(b) the half filling correlations remain long-ranged, consistent with the robustness of N\'{e}el order at moderate frustration. By $t^{\prime}=0.9$ in Fig.~\ref{fig:csdw_rp}(c), however, all three curves, including the one at half filling, decay within a single lattice spacing. Hence, the long-range order has been destroyed and only short-range correlations survive. Taken together, Figs.~\ref{fig:sharpness_combined} and~\ref{fig:csdw_rp} indicate that long-range order is confined to near half filling and small $t^{\prime}$, where $R_p$ is large, while both hole doping and increasing $t^{\prime}$ turn the long-ranged correlations into short-ranged ones. The finite-size scaling of $S^{z}(\mathbf{Q})/N$ in Appendix~B is consistent with this: it extrapolates to zero at the doped points and to a finite value at half filling.

\subsection{Magnetic correlations near and away from half filling}
\label{sec:regimes}

\begin{figure*}[!t]
    \centering
    \includegraphics[width=0.8\textwidth]{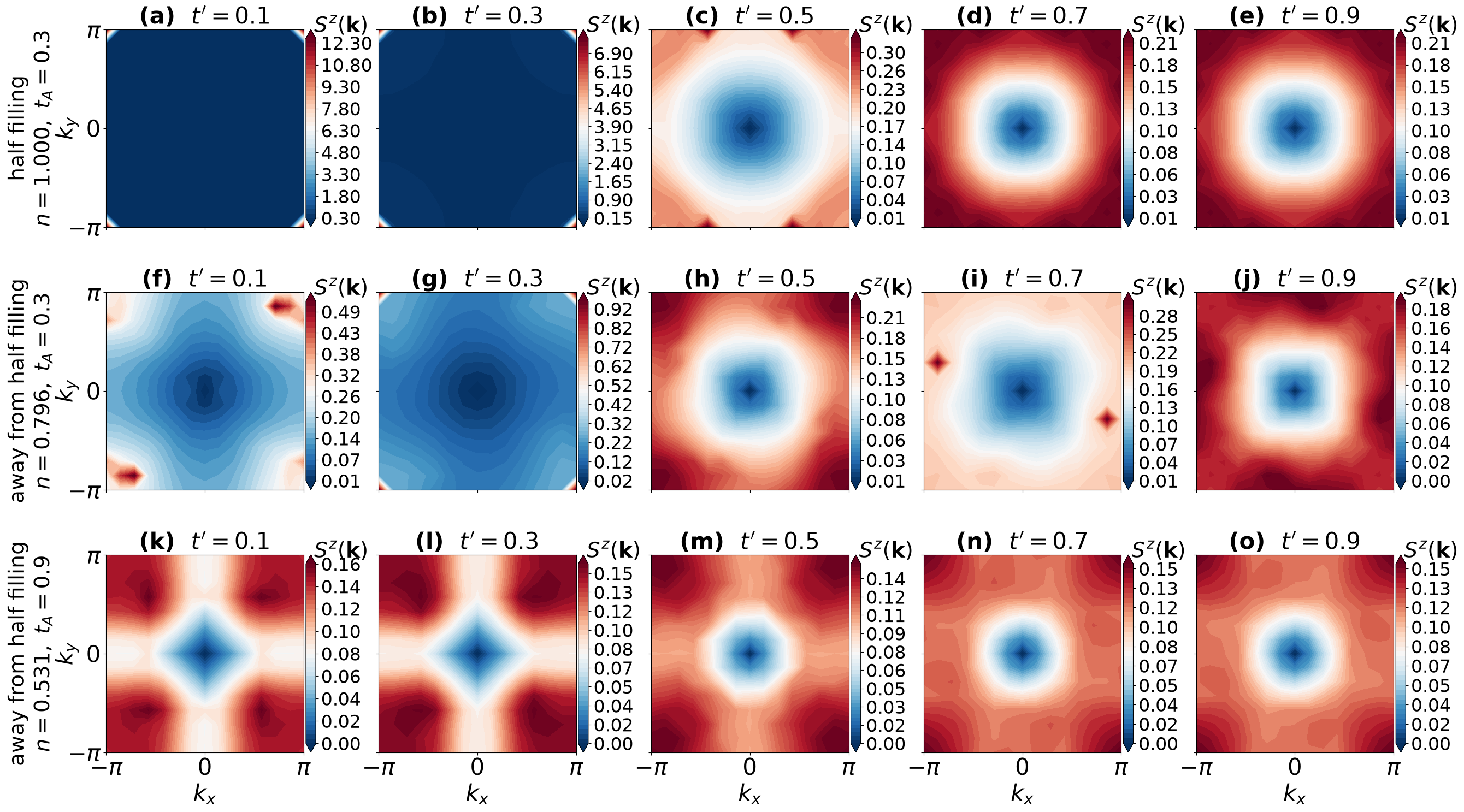}
    \caption{Spin structure factor $S^z(\mathbf{k})$ for increasing NNN hopping $t^{\prime}=0.1$, $0.3$, $0.5$, $0.7$, $0.9$. Top row (a)--(e): half filling, $n=1.000$, $t_A=0.3$. Middle row (f)--(j): away from half filling at small anisotropy, $n=0.796$, $t_A=0.3$. Bottom row (k)--(o): away from half filling at large anisotropy, $n=0.531$, $t_A=0.9$. At half filling, the sharp $(\pi,\pi)$ N\'{e}el peak at small $t^{\prime}$ collapses in magnitude and moves off the corner to a spiral wavevector as $t^{\prime}$ increases. Away from half filling at small $t_A$, the dominant peak moves from diagonal correlations $\mathbf{Q}=(q,q)$ at $t^{\prime}=0.1$, to N\'{e}el correlations $\mathbf{Q}=(\pi,\pi)$ at $t^{\prime}=0.3$, and toward stripe correlations $\mathbf{Q}=(\pi,0)/(0,\pi)$ at $t^{\prime}=0.9$. At large $t_A$, the diagonal correlations at $t^{\prime}=0.1$ are instead replaced by N\'{e}el as $t^{\prime}$ increases, and N\'{e}el persists without giving way to stripe.}
    \label{fig:sz_nearaway}
\end{figure*}

The combined phase diagram of Fig.~\ref{fig:phase_combined} organizes naturally into two regimes set by the filling, that is, by how well the Fermi surface supports scattering at the zone corner $(\pi,\pi)$. Two trends hold across all six panels and frame the discussion. First, the anisotropy $t_A$ always promotes the $(\pi,\pi)$ correlations. Increasing $t_A$ moves the dominant correlation toward N\'{e}el. Second, the NNN hopping $t^{\prime}$ works against the $(\pi,\pi)$ correlations, but in opposite ways in the two regimes, driving spiral correlations near half filling and stripe correlations away from it.

Near half filling ($n \gtrsim 0.85$), the dominant correlation is N\'{e}el for weak $t^{\prime}$. At $t^{\prime}=0$ and $0.1$ the high-filling edge is N\'{e}el for all $t_A$ in Fig.~\ref{fig:phase_combined}(a,b). The N\'{e}el region is widest at $t^{\prime}=0.3$, where it extends down to $n\approx0.7$ for almost all $t_A$ in Fig.~\ref{fig:phase_combined}(c). A moderate $t^{\prime}$ first suppresses the competing diagonal and spiral channels before it is strong enough to frustrate the $(\pi,\pi)$ correlations themselves, so the N\'{e}el region is largest at intermediate $t^{\prime}$ rather than at $t^{\prime}=0$. As $t^{\prime}$ grows further, the N\'{e}el correlations survive only at large $t_A$, while at small $t_A$ the high-filling edge turns spiral as shown in Fig.~\ref{fig:phase_combined}(d,e), and by $t^{\prime}=0.9$ the edge is spiral across almost all $t_A$, with N\'{e}el confined to large $t_A$ and intermediate filling, as seen in Fig.~\ref{fig:phase_combined}(f).

The top row of Fig.~\ref{fig:sz_nearaway} makes this explicit at half filling. At $t^{\prime}=0.1$ and $0.3$ in Fig.~\ref{fig:sz_nearaway}(a,b), $S^z(\mathbf{k})$ has a sharp, large peak at the zone corner $\mathbf{Q}=(\pi,\pi)$, the N\'{e}el state. As $t^{\prime}$ increases to $0.5$ and beyond in Fig.~\ref{fig:sz_nearaway}(c)--(e), the peak moves off the corner to a spiral wavevector and its magnitude collapses, consistent with the loss of long-range order shown in Fig.~\ref{fig:csdw_rp}. Near half filling, therefore, increasing $t^{\prime}$ converts N\'{e}el into spiral~\cite{Igoshev2010, Klett2024}, while $t_A$ delays the conversion to larger $t^{\prime}$.

Away from half filling ($n \lesssim 0.8$), the loss of $(\pi,\pi)$ nesting upon doping releases the ordering wavevector from the zone corner and exposes the competition among diagonal, spiral, and stripe correlations~\cite{Igoshev2010}. At small $t^{\prime}$, the doped region shows diagonal $(q,q)$ correlations at large $t_A$~\cite{Miyazaki2022} and spiral $(\pi,q)/(q,\pi)$ correlations at small $t_A$, with no stripe in Fig.~\ref{fig:phase_combined}(a,b). The diagonal region is fragile; by $t^{\prime}=0.3$ it survives only in the corner of large $t_A$ and low filling, as seen in Fig.~\ref{fig:phase_combined}(c). Figure~\ref{fig:phase_combined}(d)--(f) shows that, as $t^{\prime}$ increases further, stripe correlations take over the small-$t_A$ region and spread to higher filling, while N\'{e}el persists at large $t_A$.

The middle row of Fig.~\ref{fig:sz_nearaway} follows the same evolution at $n=0.796$. The dominant $S^z(\mathbf{k})$ peak moves from the zone diagonal $\mathbf{Q}=(q,q)$ at $t^{\prime}=0.1$ in Fig.~\ref{fig:sz_nearaway}(f), to the zone corner $\mathbf{Q}=(\pi,\pi)$ at $t^{\prime}=0.3$ in Fig.~\ref{fig:sz_nearaway}(g). As $t^{\prime}$ increases further, the $(\pi,\pi)$ peak weakens and its weight migrates toward the zone boundary~\cite{Mushkaev2025}, reaching the stripe wavevector $\mathbf{Q}=(\pi,0)/(0,\pi)$ at $t^{\prime}=0.9$ in Fig.~\ref{fig:sz_nearaway}(j). The bottom row of Fig.~\ref{fig:sz_nearaway}, at larger anisotropy $t_A=0.9$ and doping $n=0.531$, shows how this evolution changes when $t_A$ is strong. The diagonal correlations at $t^{\prime}=0.1$ in Fig.~\ref{fig:sz_nearaway}(k) give way to $(\pi,\pi)$ N\'{e}el correlations as $t^{\prime}$ increases, and the N\'{e}el peak persists across Fig.~\ref{fig:sz_nearaway}(l)--(o). Away from half filling, therefore, increasing $t^{\prime}$ rotates the dominant correlation from diagonal toward stripe at small $t_A$, while at large $t_A$ the anisotropy instead holds the system at N\'{e}el.

These two regimes together show how $t^{\prime}$ reshapes the phase diagram. At $t^{\prime}=0$, N\'{e}el correlations sit at high filling and are lost as the system is doped. Increasing $t^{\prime}$ changes this in two ways. The high-filling edge turns from N\'{e}el into spiral, and stripe correlations grow out of the low $t_A$, low-filling corner. As a result, the N\'{e}el region shifts away from the high-filling edge toward large $t_A$ and intermediate filling. In all cases $t_A$ favors $(\pi,\pi)$ N\'{e}el correlations and $t^{\prime}$ works against it, while the filling decides which of the competing correlations takes over. This behavior follows from the way the band structure changes with filling. The dominant correlations develop at the wavevector that best nests the spin-resolved Fermi surface, so that reshaping this surface through filling, $t_A$, or $t^{\prime}$ directly moves the preferred ordering wavevector. Near half filling the bands still favor scattering at $(\pi,\pi)$, so N\'{e}el correlations are preferred. Away from half filling the reshaped, spin-split Fermi surfaces instead favor the diagonal and zone-boundary correlations.

Setting $t_A = 0$ recovers the spin-dependent NNN altermagnetic Hubbard model~\cite{Mushtaq2026}, in which the magnetic frustration originates solely from the spin-dependent NNN hopping $t^{\prime}$. There, near-perfect $(\pi,\pi)$ nesting stabilizes N\'{e}el correlations at high filling and small $t^{\prime}$, while increasing $t^{\prime}$ or doping shifts the dominant wavevector toward spiral $\mathbf{Q}=(\pi,q)/(q,\pi)$ and, at low filling and large $t^{\prime}$, stripe $\mathbf{Q}=(\pi,0)/(0,\pi)$ correlations. This NNN limit serves as the baseline from which the NN model and the combined model emerge as modifications.

The diagonal correlations depend on both hoppings. They are almost absent in the NNN model and grow with $t_A$. Increasing $t'$ suppresses them again, and at large $t_A$ they give way to N\'eel rather than stripe correlations. This behavior is the same on all three lattice sizes, and neither model alone shows it.

\begin{figure*}[!t]
    \centering
    \includegraphics[width=0.78\textwidth]{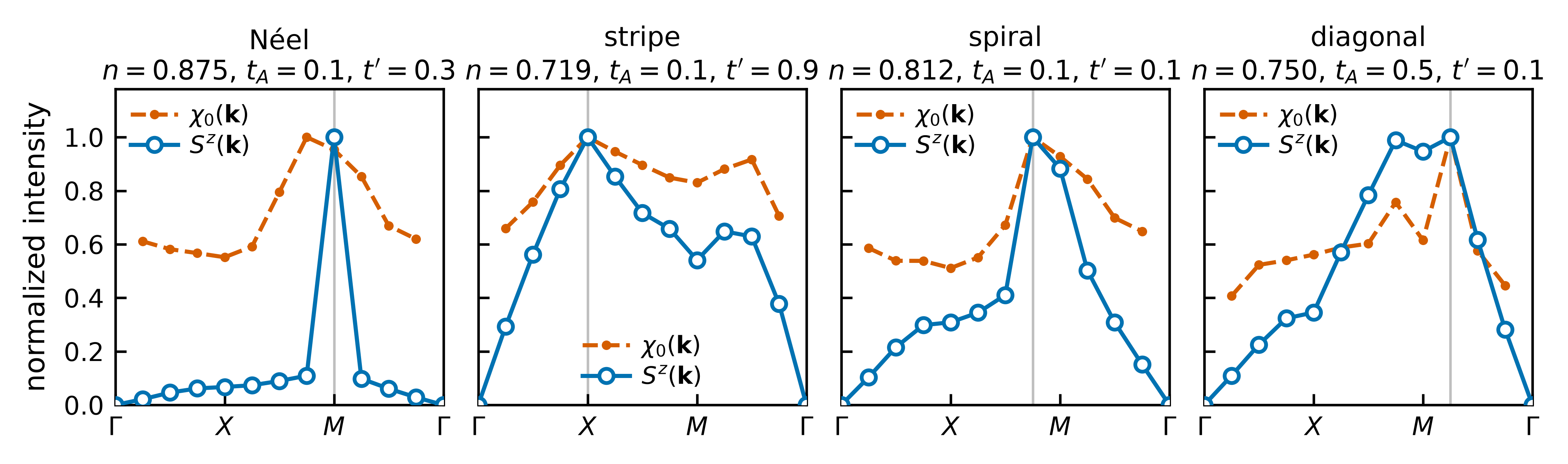}
    \caption{\label{fig:chi0}Noninteracting susceptibility $\chi_0(\mathbf{k})$ and CPQMC spin structure factor $S^{z}(\mathbf{k})$ along the path $\Gamma=(0,0)\to X=(\pi,0)\to M=(\pi,\pi)\to\Gamma$ of the $8\times8$ momentum grid at $U=4t$, for one example point of each correlation type. Each curve is divided by its largest value on the path, excluding $\Gamma$. CPQMC error bars are smaller than the symbols. The vertical line marks the $S^{z}(\mathbf{k})$ maximum. For the stripe, spiral and diagonal correlations, $S^{z}(\mathbf{k})$ peaks where $\chi_0$ is largest. For N\'eel, $\chi_0$ has only a weak maximum near $M$, while $S^{z}(\mathbf{k})$ is sharply peaked at $(\pi,\pi)$.}
\end{figure*}

To connect the band structure of Sec.~\ref{sec:noninteracting} with the interacting results, we computed the noninteracting spin susceptibility of the bands of Eqs.~(\ref{eq:disp_up}) and (\ref{eq:disp_dn}),
\begin{equation}
\chi_0(\mathbf{k})=\frac{1}{4N_p}\sum_{\sigma,\mathbf{p}}
\frac{f(\varepsilon_\sigma(\mathbf{p}))-f(\varepsilon_\sigma(\mathbf{p}+\mathbf{k}))}
{\varepsilon_\sigma(\mathbf{p}+\mathbf{k})-\varepsilon_\sigma(\mathbf{p})},
\label{eq:chi0}
\end{equation}
where $f$ is the Fermi function at $T=0.01t$ and the sum runs over $N_p$ points of a $\mathbf{p}$ grid ten times finer than the lattice. We evaluate $\chi_0$ on the same $\mathbf{k}$ grid as the simulations. Figure~\ref{fig:chi0} compares it with $S^{z}(\mathbf{k})$ along $\Gamma\to X\to M\to\Gamma$ for one example point of each correlation type. For the stripe, spiral and diagonal correlations, $S^{z}(\mathbf{k})$ peaks where $\chi_0$ is largest, and the two peaks have similar widths, so these correlations follow the nesting of the spin-split Fermi surfaces. In the N\'eel case $\chi_0$ is also largest around $(\pi,\pi)$, but its maximum is broad. On the full $L=14$ grid, $\chi_0$ and $S^{z}(\mathbf{k})$ have their maximum at the same wavevector at most points for $t'\le0.1$, at a minority of points for $t'\ge0.3$, and almost never at the N\'eel points. Nesting therefore selects the dominant wavevector at small $t'$. At larger $t'$ the interaction selects it, in particular for the N\'eel correlations.

\begin{figure*}[!t]
\includegraphics[width=0.8\textwidth]{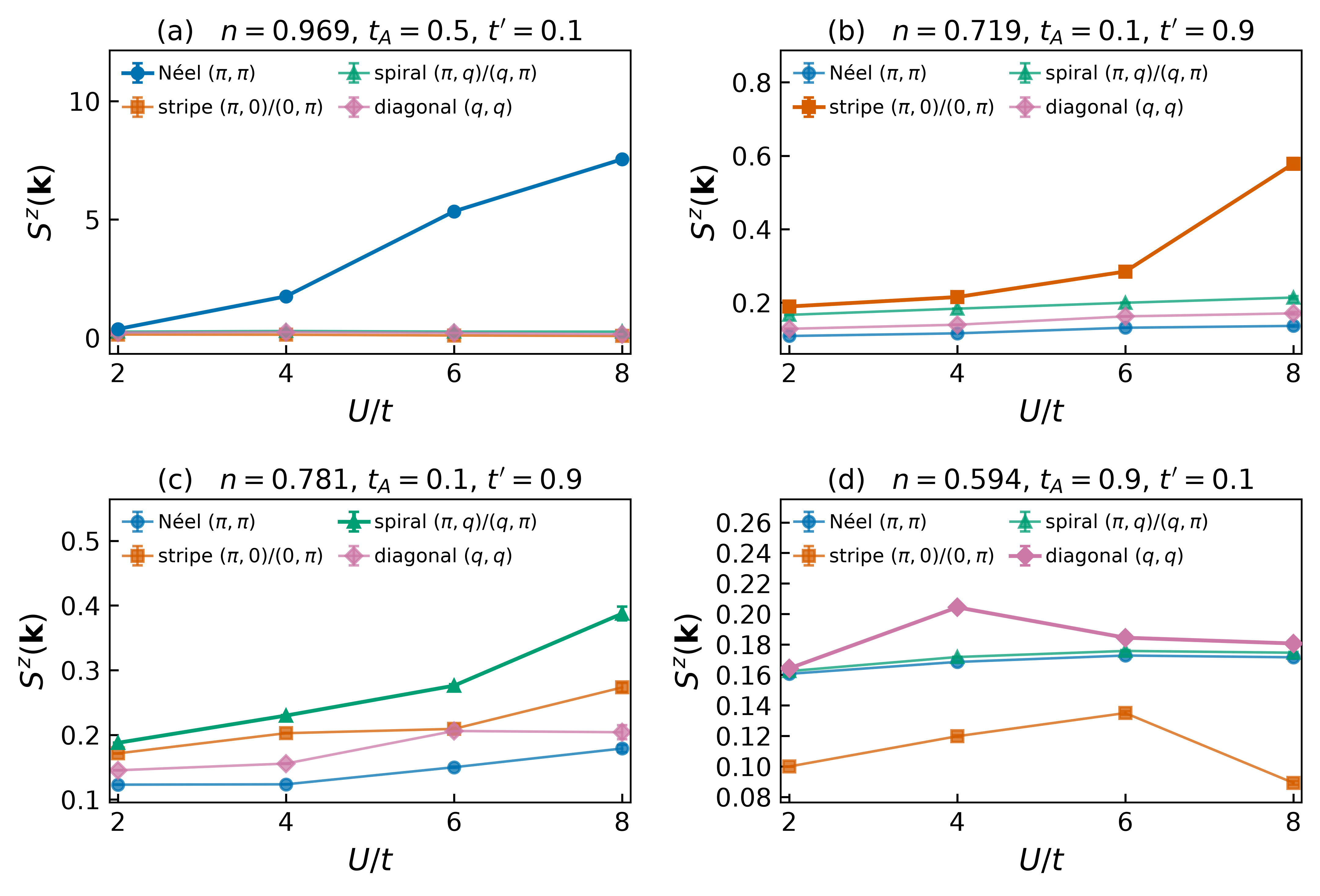}
\caption{\label{fig:U}Dependence on $U$ of the spin structure factor on the $L=8$ lattice, at four representative points, one for each correlation type. Each panel shows $S^{z}(\mathbf{k})$ at the four characteristic wavevectors. Error bars are smaller than the symbols where not visible. (a) Near half filling the N\'eel peak grows by more than an order of magnitude between $U=2t$ and $8t$, while the other three stay flat. (b) and (c) In the doped region the stripe and spiral peaks both grow with $U$. In (b) the stripe peak grows fastest and pulls away above $U=6t$; in (c) the spiral peak stays the largest. (d) At large $t_A$ and small $t'$ the diagonal peak leads at every $U$ but hardly changes, so this correlation does not grow with the interaction.}
\end{figure*}

\begin{figure*}[!t]
    \centering
    \includegraphics[width=0.8\textwidth]{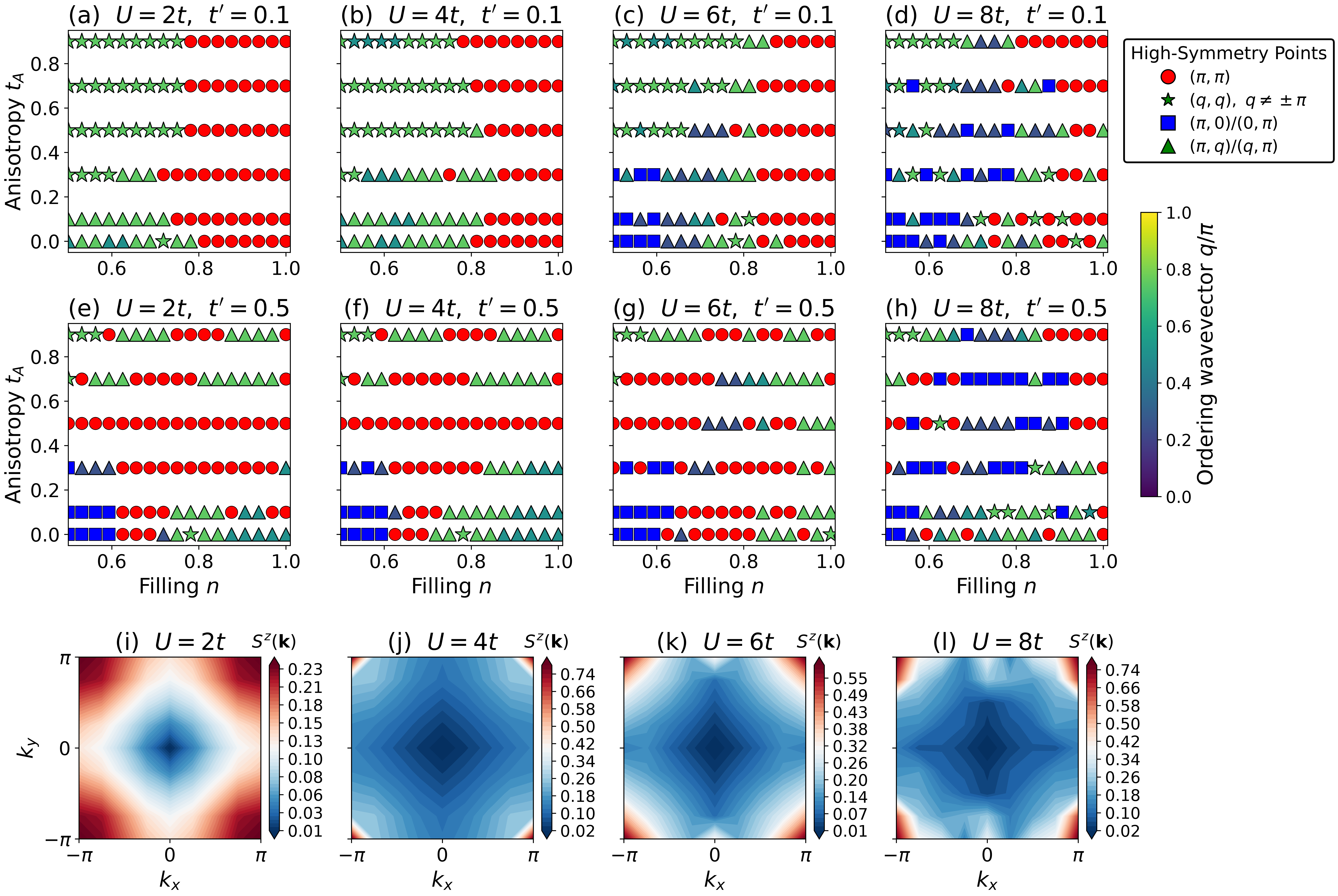}
    \caption{\label{fig:umaps}(a)--(h) Magnetic phase diagram on the $L=8$ lattice at $U=2t$, $4t$, $6t$ and $8t$ for $t'=0.1$ and $t'=0.5$. (i)--(l) Spin structure factor $S^{z}(\mathbf{k})$ over the Brillouin zone at $n=0.875$, $t_A=0.9$ and $t'=0.1$ for the same four values of $U$. The peak sits at the N\'eel wavevector $(\pi,\pi)$ at all four values of $U$. It is weak and broad at $U=2t$ and is sharp and much stronger from $U=4t$ on.}
\end{figure*}

All results above are at $U=4t$. To see how they depend on the interaction, we repeated the $(n,t_A)$ scan on the $L=8$ lattice at $U=2t$, $6t$ and $8t$, for $t'=0.1$, $0.5$ and $0.9$. Figure~\ref{fig:U} follows one representative point of each correlation type. Near half filling the N\'eel peak grows by more than an order of magnitude over this range of $U$, while the stripe, spiral and diagonal peaks at the same point stay flat, so the $(\pi,\pi)$ tendency strengthens toward the Heisenberg limit. In the doped region the stripe and spiral peaks grow more slowly. At the stripe point the stripe peak grows fastest and pulls away from the others between $U=6t$ and $8t$, while at the spiral point the spiral peak stays the largest. At large $t_A$ and small $t'$ the diagonal correlations dominate but barely change with $U$, which indicates that the diagonal correlations follow the band anisotropy rather than the interaction. The phase diagrams in Fig.~\ref{fig:umaps} change gradually with $U$ over most of the plane. The bottom row of Fig.~\ref{fig:umaps} shows the structure factor behind one point of the N\'eel region. The peak sits at $(\pi,\pi)$ at every $U$; it is weak and broad at $U=2t$ and is sharp and much stronger from $U=4t$ on. The diagonal region still grows with $t_A$ at $U=6t$ but no longer at $U=8t$, so the joint effect of $t_A$ and $t'$ belongs to weak-to-intermediate coupling.

\section{Conclusion}

In this work, we have mapped the magnetic phase diagram of the square-lattice Hubbard model at $U=4t$ as a function of electron filling $n$, NN hopping anisotropy $t_A$, and NNN hopping $t'$. Using CPQMC with a Hartree--Fock trial wave function, we identified four distinct magnetic correlations and tracked their boundaries across a broad region of the $(n,t',t_A)$ parameter space. These correlations are N\'{e}el $\mathbf{Q}=(\pi,\pi)$, stripe $\mathbf{Q}=(\pi,0)/(0,\pi)$, spiral $\mathbf{Q}=(\pi,q)/(q,\pi)$, and diagonal $\mathbf{Q}=(q,q)$. The $R_p$ sharpness index introduced here provides a quantitative measure of magnetic correlation quality that does not rely on prior knowledge of the ordering wavevector, offering a tool that remains useful in regimes where multiple wavevectors compete.

A key finding of this study is that $t_A$ acts as a symmetry-breaking field in momentum space. In the absence of hopping anisotropy, the square-lattice Fermi surface respects the full $C_4$ point-group symmetry, and SDW instabilities at $\mathbf{Q}=(\pi,0)$ and $\mathbf{Q}=(0,\pi)$ are degenerate. Introducing $t_A$ differentiates the spin-up and spin-down dispersions, shifting their respective van Hove singularities. This shift preferentially weights the spin structure factor along the $(q,q)$ direction, stabilizing the diagonal correlation over a sizable region of filling. This diagonal correlation is driven by the spin-dependent anisotropy $t_A$ and vanishes as $t_A\to0$, distinguishing it from the spin-independent mechanism operative in the conventional $t$--$t'$ Hubbard model. The diagonal region is therefore a sign of spin-dependent hopping anisotropy, and it could be tested in cold-atom systems where $t_A\gtrsim0.5$ can be engineered. The role of $t'$ is complementary. Finite $t'$ introduces frustration that weakens perfect $(\pi,\pi)$ nesting, suppressing N\'{e}el correlations away from half filling and expanding the region occupied by spiral and stripe correlations. Three features of this phase diagram do not appear in either the NN model or the NNN model alone. The diagonal correlations of the NN model are suppressed by $t'$ and replaced by N\'eel and spiral correlations, whereas the NNN model gives stripe correlations there. N\'eel correlations survive at large $t_A$ and large $t'$, where the NNN model alone gives stripe or spiral correlations. At fixed $n$ and $t_A$, increasing $t'$ changes the dominant correlation from diagonal to N\'eel to stripe (Fig.~\ref{fig:sz_nearaway}), a sequence that requires both hoppings. Ultracold fermions in optical lattices are a natural place to test these results, since the hopping anisotropy, the interaction and the filling can all be tuned independently over the range studied here~\cite{Das2024}. The sequence of diagonal, spiral and stripe correlations with increasing $t'$ is the clearest prediction to look for.

\begin{figure*}[!t]
  \centering
  \includegraphics[width=0.6\textwidth]{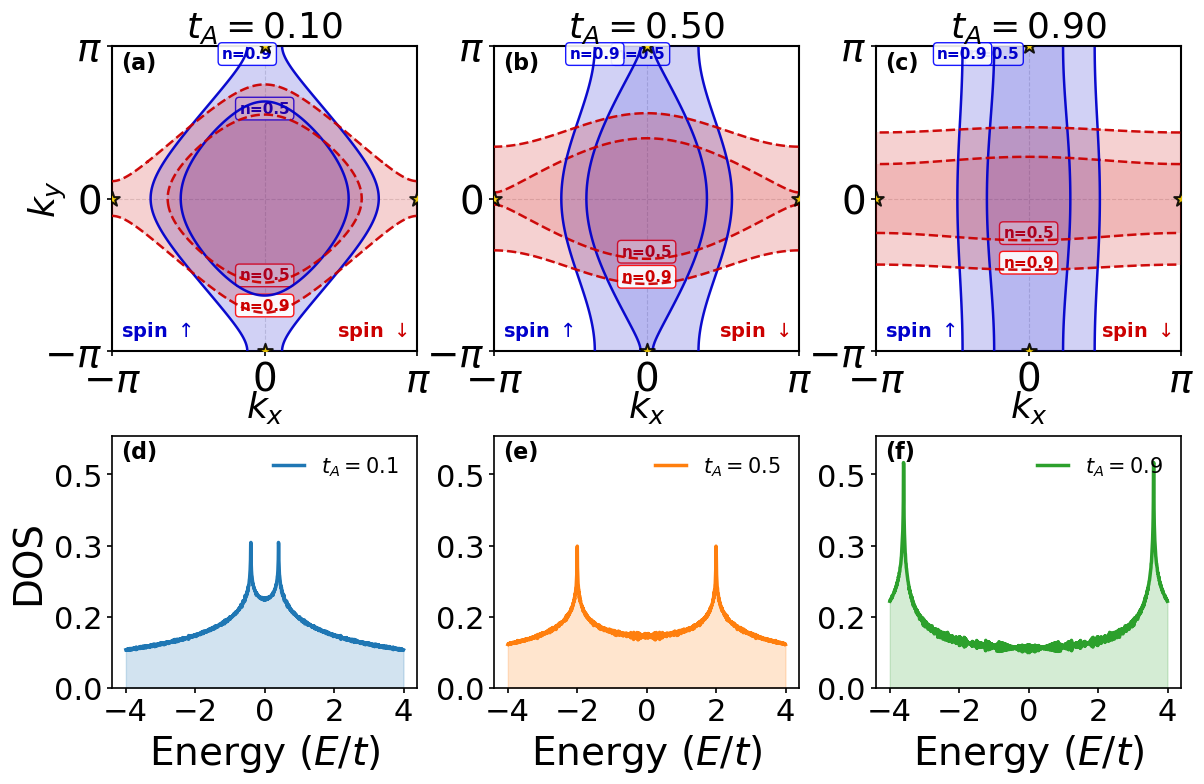}
  \caption{Noninteracting Fermi surfaces and density of states for the anisotropic NN model. Panels~(a)--(c) show spin-$\uparrow$ (blue, solid) and spin-$\downarrow$ (red, dashed) Fermi contours for $n=0.5$ and $n=0.9$ at three values of $t_A$. Gold stars mark the van~Hove saddle points at $(\pm\pi,0)$ and $(0,\pm\pi)$. Panels~(d)--(f) show the total DOS computed from spin sectors. The van~Hove peak splitting $\pm 4t_A$ grows with $t_A$ and tracks the Fermi-surface topology change shown above.}
  \label{fig:fs_dos}
\end{figure*}

In summary, we show that the combination of NN hopping anisotropy and NNN hopping produces four distinct, parameter-dependent magnetic correlations in the square-lattice Hubbard model. The three control parameters $n$, $t'$, and $t_A$ cooperate to generate a magnetic-correlation regime that cannot be understood by varying any single parameter in isolation. At small $t'$ the nesting of the spin-split Fermi surfaces explains the dominant wavevectors; at larger $t'$ the interaction sets them. Future work should address finite-size scaling of the phase boundaries with system sizes beyond $L=14$ and the relationship between the magnetic phases identified here and the symmetry of any superconducting state that emerges upon suppression of magnetic order by additional doping or pressure.

\section*{Acknowledgment}
This work is supported by the Shenzhen Fundamental Research Program (No.~JCYJ20250604145655074), and Shenzhen Key Laboratory of Advanced Functional Carbon Materials Research and Comprehensive Application (Grant No. ZDSYS20220527171407017).

\begin{figure*}[!t]
    \centering
    \includegraphics[width=0.8\textwidth]{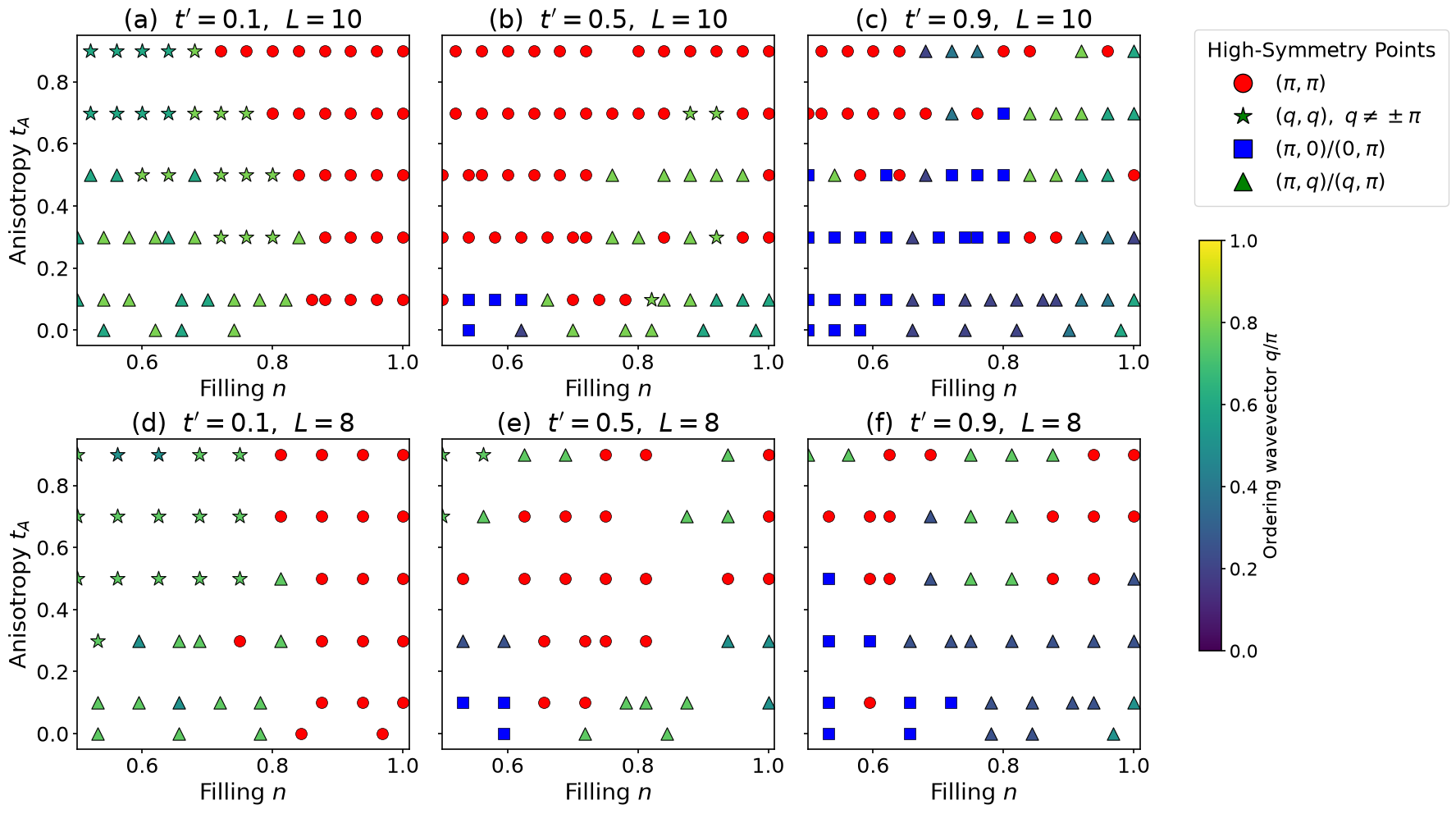}
    \caption{Magnetic phase diagram of the combined altermagnetic Hubbard model ($U = 4t$) in the $(n, t_A)$ plane for $L = 10$ panels~(a)--(c) and $L = 8$ panels~(d)--(f), at (a,d)~$t^{\prime} = 0.1$, (b,e)~$t^{\prime} = 0.5$, and (c,f)~$t^{\prime} = 0.9$. The dominant $S^z(\mathbf{k})$ peak position classifies each point as N\'{e}el correlations $\mathbf{Q} = (\pi, \pi)$, spiral correlations $\mathbf{Q} = (\pi, q)/(q, \pi)$, stripe correlations $\mathbf{Q} = (\pi, 0)/(0, \pi)$, or diagonal correlations $\mathbf{Q} = (q, q)$ with $q \neq \pm\pi$. The colorbar encodes the ordering wavevector $q/\pi$. The qualitative phase structure and boundaries are consistent with the $L = 14$ results in Fig.~\ref{fig:phase_combined}, confirming the robustness of the identified magnetic correlations against finite-size effects.}
    \label{fig:phase_L8}
\end{figure*}

\appendix
\section*{Appendix A: Noninteracting electronic structure of the NN model}
\label{app:noninteracting}

How spin-dependent NN hopping reconstructs the noninteracting electronic structure is shown in Figure~\ref{fig:fs_dos}. For $t_A=0.1$, the spin-$\uparrow$ and spin-$\downarrow$ Fermi contours remain close to the usual square-lattice form, with only moderate anisotropic distortion. With increasing $t_A$, however, the two spin sectors are stretched in opposite directions and the contours evolve into open quasi-one-dimensional sheets. Increasing $t_A$ thus drives a clear topology change of the spin-resolved low-energy structure.

Filling is crucial to this evolution. The Fermi contours stay closed up to large $t_A$ at $n=0.5$, but they open up at smaller anisotropy at $n=0.9$. Because the Fermi level more efficiently samples the anisotropic part of the dispersion at higher filling, the effect of spin-dependent hopping is more pronounced. This filling dependence is significant because it directly modifies the available low-energy scattering channels that subsequently define the dominant magnetic response. The DOS provides a complementary picture. The van~Hove feature splits into two peaks and the low-energy DOS at $E=0$ decreases with increasing $t_A$. These two peaks arise because spin-dependent anisotropy shifts the saddle-point energies in opposite directions while keeping them stationary at $(\pi,0)$ and $(0,\pi)$ in momentum space. At $t_A=0$, the van~Hove singularities of the two spin species are degenerate at $E=0$, producing a single central peak in the total DOS. For finite $t_A$, this degeneracy is lifted and the saddle-point energies shift to $\pm 4t_A$, so the total DOS develops two symmetric van~Hove peaks~\cite{Rao2025,Parthenios2025,Markiewicz1997,Irkhin2001,YuanIsobeFu2019}. The DOS thus provides a complementary signature that finite $t_A$ reorganizes the low-energy phase space before interactions are included.

\begin{figure*}[p]
    \centering
    \includegraphics[width=0.66\textwidth]{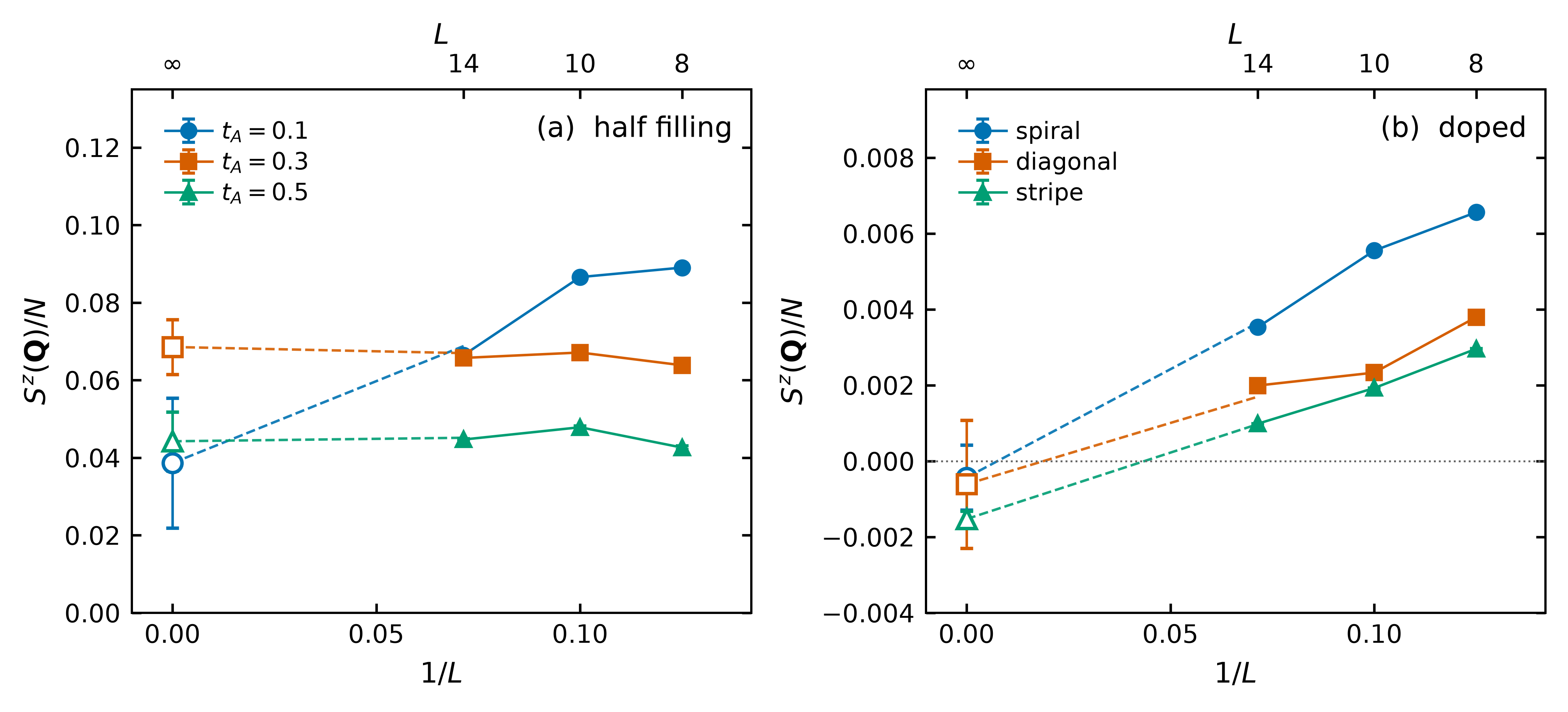}
    \caption{\label{fig:fss} Finite-size scaling of $S^{z}(\mathbf{Q})/N$ at $U=4t$ for $L=8$, $10$ and $14$. Filled symbols are the simulated lattices, dashed lines are weighted straight-line fits in $1/L$, and open symbols are the extrapolated $L\to\infty$ values. (a) Half filling, $n=1$, $t'=0.1$: the extrapolated values are finite for $t_A=0.3$ and $0.5$, consistent with long-range N\'eel order. (b) Doped points at $(t_A,t')=(0.3,0.1)$ (spiral), $(0.7,0.1)$ (diagonal) and $(0.3,0.9)$ (stripe), at the closed-shell filling closest to $n=0.84$, $0.72$ and $0.77$ on each lattice: the extrapolated values are zero within error for all three correlation types. On each lattice $\mathbf{Q}$ is the position of the largest $S^{z}(\mathbf{k})$, and it belongs to the same correlation type on all three lattices.}

\vspace{4mm}
    \centering
    \includegraphics[width=0.66\textwidth]{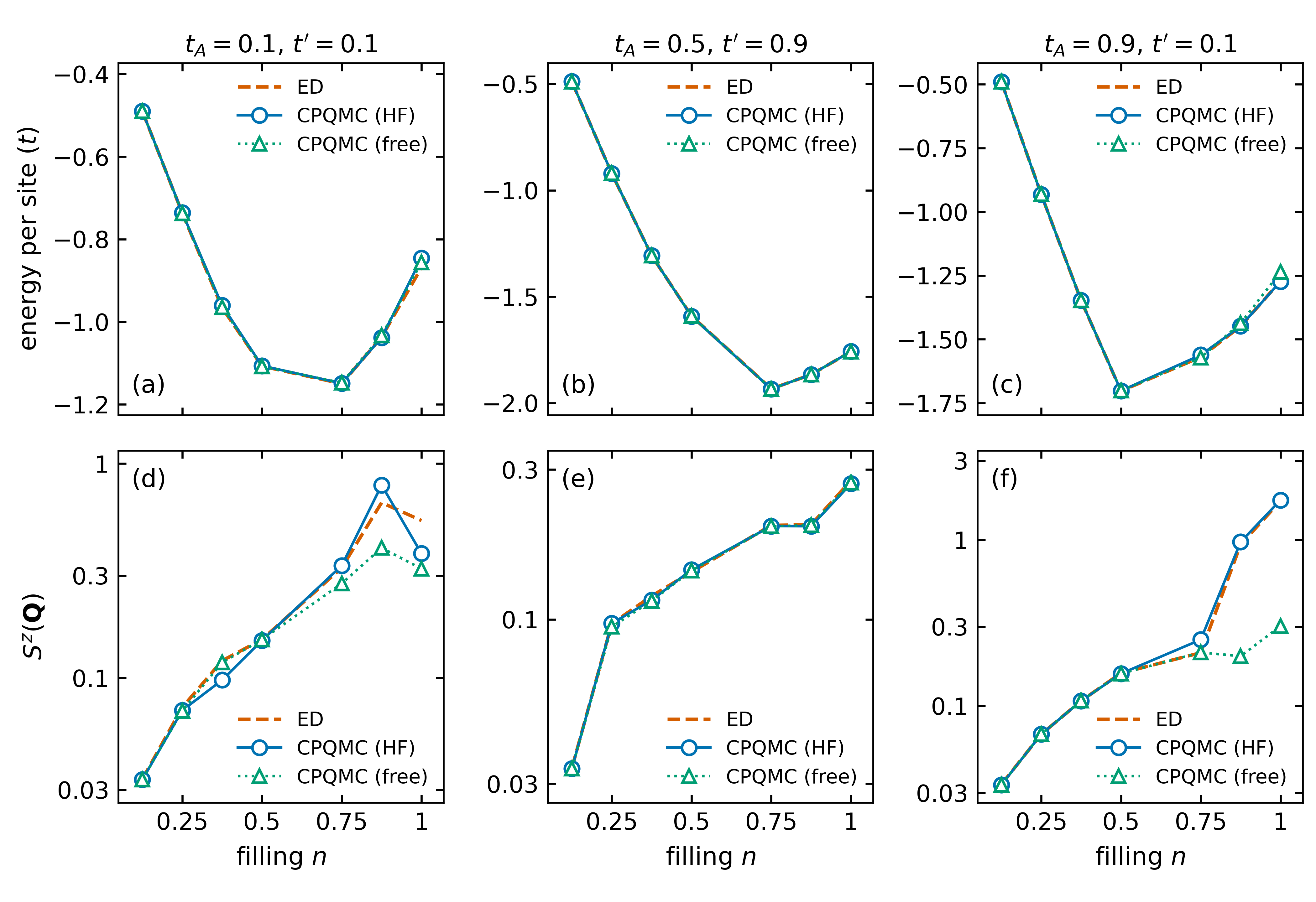}
    \caption{\label{fig:ed}ED benchmark of CPQMC on a $4\times4$ cluster at $U=4t$ for three hopping sets $(t_A,t')$, at every filling accessible on the cluster. Top row: total energy per site. Bottom row: $S^{z}(\mathbf{Q})$ at the wavevector of the ED maximum, on a logarithmic scale. Orange dashed line: exact diagonalization; blue circles: CPQMC with the unrestricted Hartree--Fock trial wave function; green triangles: CPQMC with the free-electron trial wave function ($\Delta\tau=0.01$, $N_w=1000$). CPQMC error bars are smaller than the symbol size.}

\vspace{5mm}
\makeatletter\def\@captype{table}\makeatother

\caption{\label{tab:wf}Trial-wave-function dependence at eight test points on the $L=8$ lattice at $U=4t$. Here $m_{\mathrm{loc}}=\bigl[N^{-1}\sum_i\langle n_{i\uparrow}-n_{i\downarrow}\rangle^{2}\bigr]^{1/2}$ is the root-mean-square local magnetization of the main Hartree--Fock trial state; it is zero for a paramagnet and finite for any spin-density wave. The other columns give the CPQMC peak height $S^{z}(\mathbf{Q})$ and its correlation type (N N\'eel, St stripe, Sp spiral, D diagonal) for five trial states: the main Hartree--Fock state (HF), the free-electron determinant (free), and Hartree--Fock states started from a N\'eel, stripe or diagonal spin density.}
\begin{ruledtabular}
\begin{tabular}{cccclllll}
$n$ & $t_A$ & $t'$ & $m_{\mathrm{loc}}$ & HF & free & N\'eel seed & stripe seed & diagonal seed \\
\hline
0.562 & 0.9 & 0.5 & 0.010 & 0.167 D & 0.168 D & 0.168 D & 0.168 D & 0.168 D \\
0.594 & 0.1 & 0.5 & 0.025 & 0.170 St & 0.170 St & 0.169 St & 0.170 St & 0.169 St \\
0.719 & 0.1 & 0.9 & 0.004 & 0.215 St & 0.214 St & 0.215 St & 0.215 St & 0.215 St \\
0.750 & 0.5 & 0.1 & 0.253 & 0.444 D & 0.280 D & 0.283 D & 0.283 D & 0.283 D \\
0.781 & 0.0 & 0.1 & 0.385 & 0.457 Sp & 0.302 Sp & 0.558 N & 0.305 Sp & 0.558 N \\
1.000 & 0.1 & 0.1 & 0.684 & 5.707 N & 1.367 N & 5.709 N & 1.492 N & 5.711 N \\
1.000 & 0.1 & 0.5 & 0.474 & 1.111 Sp & 0.252 N & 2.961 N & 0.257 St & 0.358 Sp \\
1.000 & 0.5 & 0.5 & 0.003 & 0.301 N & 0.298 N & 0.298 N & 0.296 N & 0.301 N \\
\end{tabular}
\end{ruledtabular}

\end{figure*}

\section*{Appendix B: Finite-size dependence of the combined model phase diagram}
\label{app:finite_size}

We compare the dominant $S^z(\mathbf{k})$ peak locations across the $(n, t_A)$ plane for $L = 8$ and $L = 10$ to assess finite-size effects in the combined altermagnetic model. The overall distribution of the peak locations is robust across all three system sizes studied. The resulting phase diagrams are shown in Fig.~\ref{fig:phase_L8}. All four magnetic correlations, namely N\'{e}el correlations at $\mathbf{Q} = (\pi, \pi)$, stripe correlations at $\mathbf{Q} = (\pi, 0)/(0, \pi)$, spiral correlations at $\mathbf{Q} = (\pi, q)/(q, \pi)$, and diagonal correlations at $\mathbf{Q} = (q, q)$ with $q \neq \pm\pi$, are reproduced at $L = 8$ and $L = 10$, and the systematic evolution of phase boundaries with $t_A$ is consistently preserved.

The primary source of finite-size effects is the discretization of the Brillouin zone. The N\'{e}el and stripe correlations peak at zone high-symmetry points present on any grid and are therefore exactly resolved at all system sizes. The spiral and diagonal regions, whose ordering wavevectors vary continuously with filling and hopping parameters, are most sensitive to momentum resolution, and points near phase boundaries may be reclassified as $L$ increases. Despite these local variations, no correlation type appears or disappears between $L = 8$, $L = 10$, and $L = 14$, confirming the robustness of the phase structure reported in the main text.

To separate long-range order from short-range correlations, we also studied how $S^{z}(\mathbf{Q})/N$ changes with lattice size (Fig.~\ref{fig:fss}). For $L\to\infty$ it goes to the square of the order parameter when long-range order is present, and to zero otherwise. At half filling the extrapolated values are finite at $t_A=0.3$ and $0.5$, consistent with long-range N\'eel order. At the doped points they are zero within error for all three correlation types. The half-filling values, $0.04$--$0.07$, are close to the squared staggered magnetization of the isotropic model at $U=4t$, $M_0^2=0.056$, from numerically exact quantum Monte Carlo~\cite{Qin2016}. We therefore describe the stripe, spiral and diagonal regimes as magnetic correlations, not long-range order. With only three lattice sizes, these results are indicative rather than a precise extrapolation.

\section*{Appendix C: Exact diagonalization}
\label{app:ed}

We compare CPQMC with ED on a $4\times4$ cluster at $U=4t$, using both the Hartree--Fock and the free-electron trial wave functions. The test points are the three hopping sets of Fig.~\ref{fig:ed} at every filling of the cluster, plus three more sets at $n=0.625$ and $n=1$. The median energy deviation from ED is $0.06\%$, and the largest deviations, $2.6\%$ and $2.8\%$, are at half filling. The dominant wavevector agrees with ED at nearly every test point for both trial states. Away from half filling, $S^{z}(\mathbf{Q})$ at the ED maximum agrees within about $1\%$ at most points. The largest deviations are at $t_A=0.9$ and low filling, where two peaks are nearly equal. At half filling the CPQMC value of $S^{z}(\pi,\pi)$ follows the trial state: it ranges from $30\%$ too small to four times too large with the Hartree--Fock state, and from $82\%$ too small to $7\%$ too large with the free-electron state. The exact ground state of a $4\times4$ cluster has no broken symmetry, and neither determinant reproduces it there. The benchmark therefore supports the phase assignments away from half filling, while the half-filling structure factor rests on Appendix~D. The back-propagation length does not cause this. Increasing it from 10 to 80 steps lowers $S^{z}(\pi,\pi)$ at the $4\times4$ point $(t_A,t')=(0.5,0.1)$ only from 1.64 to 1.43, compared with the exact 0.36. At the three $L=8$ points of Fig.~\ref{fig:fss}, doubling it from 40 to 80 steps lowers $S^{z}(\mathbf{Q})$ by $7$--$10\%$ without changing the dominant wavevector. The absolute values of $S^{z}(\mathbf{Q})$ in this work therefore carry a systematic uncertainty of about $10\%$.

\section*{Appendix D: Dependence on the trial wave function}
\label{app:twf}

The constrained-path approximation depends on the trial wave function. We therefore repeated eight runs on the $L=8$ lattice with four other trial states (Table~\ref{tab:wf}). The main Hartree--Fock state and the free-electron determinant give the same dominant wavevector at all but one of the test points, including all doped points. Trial states started from a N\'eel, stripe or diagonal pattern reproduce this result where one correlation type dominates. Where two types have nearly equal peak heights, the label can change: at $n=0.781$, $t_A=0$, $t'=0.1$ the N\'eel-started state gives N\'eel instead of spiral. Such points lie on the boundaries of Fig.~\ref{fig:phase_combined}, which are uncertain by one grid point in $n$ (Sec.~\ref{sec:combined_phase}). At half filling and small $t'$ the main state has N\'eel order and gives $S^{z}(\pi,\pi)$ about four times larger than the free-electron determinant. With the free-electron determinant, the finite-size extrapolation of Appendix~B falls to the noise level. The $4\times4$ benchmark of Appendix~C shows that on a small cluster neither value is exact. We keep the main Hartree--Fock state for three reasons. It is the mean-field solution at the parameters of each run. It agrees with the free-electron determinant at every doped point. Its half-filling extrapolations (Appendix~B) are close to the squared staggered magnetization of the isotropic model.

\FloatBarrier

\bibliography{references}
\end{document}